\documentclass{ptephy}

\begin{document}

\title{Efficient Acceleration of Relativistic Magnetohydrodynamic Jets}

\author{\name{Kenji Toma}{1} and \name{Fumio Takahara}{1}} 

\address{\affil{1}{Department of Earth and Space Science, Graduate School of Science, Osaka University, Toyonaka 560-0043, Japan}
\email{toma@vega.ess.sci.osaka-u.ac.jp}}

\begin{abstract}%
Relativistic jets in active galactic nuclei, galactic microquasars, and gamma-ray bursts are 
widely considered to be magnetohydrodynamically driven by black hole accretion systems,
although conversion mechanism from Poynting into particle kinetic energy flux is still open.
Recent detailed numerical and analytical studies of global structures of steady, axisymmetric 
magnetohydrodynamic (MHD) flows with specific boundary conditions have not reproduced 
as rapid an energy conversion as required by observations. In order to find more
suitable boundary conditions, we focus on the flow along a poloidal magnetic
field line just inside the external boundary, without treating transfield force balance in detail. 
We find some examples of the poloidal field structure and corresponding external pressure
profile for an efficient and rapid energy conversion as required by observations, 
and that the rapid acceleration requires a rapid decrease of the external pressure above
the accretion disk. 
We also clarify the differences between the fast magnetosonic point of the MHD flow 
and the sonic point of de Laval nozzle.
\end{abstract}

\subjectindex{xxxx, xxx}

\maketitle

\section{Introduction}
\label{sec:intro}

Collimated outflows, or jets, with relativistic speeds
are observed in active galactic nuclei (AGNs) and galactic microquasars, 
and are presumably driven in gamma-ray bursts (GRBs). A widely
discussed model for these is that rotational energies of an accretion flow around a black hole (BH) 
and/or of a BH itself are magnetohydrodynamically converted
into an outflow energy \cite{lovelace76,blandford77}, in analogy to pulsar winds \cite{goldreich69}. 
(Alternative possibility that thermal energy of a BH accretion flow is transferred into an outflow
is also discussed \cite[e.g.,][]{asano09,becker11,toma12}.) In relativistic jets driven 
magnetohydrodynamically, Poynting flux dominates the total energy output at the
launching site, while observations of AGN jets strongly suggest that the energy is dominated 
by particle kinetic energy at the emission site \cite[e.g.,][]{inoue96,kino02,sikora05}. 
It has been actively debated how Poynting flux can be efficiently converted into kinetic energy flux.

In oder to discuss the energy conversion, ideal magnetohydrodynamics (MHD) is the 
simplest approximation. We consider steady, axisymmetric outflows as further approximation.\footnote{
Unsteadiness and/or non-ideal MHD effects can be important for energy conversion
from Poynting into kinetic energy flux \cite[e.g.,][]{kirk11,granot11,mckinney11}.
}
In this context, the poloidal velocity of the flow has to exceed fast magnetosonic speed 
(i.e., pass the fast magnetosonic point)
to continue accelerating towards infinity \cite{camenzind86}. At the fast magnetosonic 
point (or the fast point), the bulk Lorentz factor $\Gamma$ becomes $\sim {\mathcal E}^{1/3}$
for ${\mathcal E} \gg 1$, where 
${\mathcal E}$ is the total energy flux per unit rest energy flux 
(more precisely, see Eq. \ref{eq:gamma_f} below) \cite{camenzind86,michel69,begelman94}. 
Thus a major energy conversion
from Poynting into kinetic energy flux, up to $\Gamma \approx {\mathcal E}$, should occur 
beyond the fast point. 
Self-similar analysis of full MHD equations including the transfield force balance has shown
that an efficient acceleration beyond the fast point can occur but that a
major energy conversion requires very long distance \cite{li92, vlahakis03}.

Pioneering works by Komissarov et al. (2007, 2009) 
\cite{komissarov07,komissarov09} have performed numerical simulations of time-dependent,
axisymmetric MHD equations over much extended region, obtaining steady solutions of 
relativistic jets with specific boundary conditions \cite[see also][]{tchekhovskoy09}. 
They show that the outflows which are
confined by a paraboloidal wall accelerate efficiently. In those outflows, poloidal magnetic
fields near the jet axis are self-collimated, leading to a decrease of the field strength
and an efficient acceleration in the main body of the flow. This result is supported by some
analytical treatments \cite{lyubarsky09,lyubarsky10,beskin06}. In particular, 
Lyubarsky (2009, 2010) \cite{lyubarsky09,lyubarsky10} constructed asymptotic solutions of 
the outflow structure far beyond the light cylinder, confined by external pressure decaying as 
a power law of height, under the
assumption that Poynting flux is not significantly converted into kinetic energy flux near 
the fast point.

These studies show a rapid energy conversion up to the equipartition level, i.e., 
the ratio of Poynting to kinetic energy flux $\sigma \sim 1$, but after that, 
the conversion rate dramatically decreases. Then the distance where $\sigma \lesssim 0.1$
is realized is too large, compared with the observational suggestions that
$\sigma \lesssim 0.1$ at the distance of the emission site $\sim 10^3 - 10^4\;r_{\rm s}$, where 
$r_{\rm s}$ is the Schwarzschild radius of a central BH \cite{inoue96,kino02,sikora05}.

However, the previous studies \cite{komissarov07,komissarov09,tchekhovskoy09,lyubarsky09,lyubarsky10}
have considered only special types of boundary conditions. Thus it may be useful to clarify
whether there can be boundary conditions and magnetic field configurations leading to 
more efficient acceleration that is compatible to the observational suggestions.
Fendt \& Ouyed (2004) \cite{fendt04} have raised a hint for this problem by only solving
the poloidal direction of motion without treating the transfield force balance \cite[see also][]{takahashi98}. 
They show that a rapid energy conversion up to $\Gamma \approx \mathcal{E}$ near the fast point occurs 
if the poloidal field strength evolves along a field line as a power law of cylindrical radius like
$B_p \propto r^{-2-q}$ with $q>0$, although the corresponding structure of the 
magnetic field was not specified.
In this paper, we investigate magnetic field structure required for a rapid major conversion of energy.
We focus on the fluid motion along a poloidal field line just inside the external boundary 
(see Figure \ref{fig:field_shape} below) and show some examples of the field line shape and 
the external pressure profile for a rapid energy conversion (Section \ref{sec:main}).
We do not treat the transfield force balance in detail.
Before showing our findings, we review basic theory of steady, axisymmetric, relativistic MHD flow (Section
\ref{sec:basic}) and discuss properties of the fast point a bit more deeply than previous papers
(Section \ref{sec:nozzle}).

\section{Basic Equations}
\label{sec:basic}

We make a brief review of theory of steady, axisymmetric, relativistic ideal-MHD flow
\cite{mestel61,bekenstein78,camenzind86,vlahakis03,lyubarsky09,beskin09}. We also assume
that the flow is cold, and neglect effects of gravitational field, for simplicity.
Basic equations are then equation of motion
\begin{equation}
\rho c^2 ({\mathbf u} \cdot \nabla) {\mathbf u} = \frac{1}{4\pi} \left[
(\nabla \cdot \mathbf{E})\mathbf{E} + (\nabla\times \mathbf{B})\times\mathbf{B} \right],
\label{eq:eom}
\end{equation}
equation of continuity
\begin{equation}
\nabla \cdot (\rho {\mathbf u}) = 0,
\label{eq:eoc}
\end{equation}
Maxwell equations
\begin{eqnarray}
\nabla \cdot {\mathbf B} = 0, \label{eq:maxwell_B}\\
\nabla \times {\mathbf E} = 0,
\label{eq:maxwell_E}
\end{eqnarray}
and ideal MHD condition
\begin{equation}
{\mathbf E} + \frac{1}{c} {\mathbf v} \times {\mathbf B} = 0,
\label{eq:mhd}
\end{equation}
where ${\mathbf u} = \Gamma {\mathbf v}/c$, and $\Gamma$ is Lorentz factor of the flow.

In the axisymmetric configuration, it is convenient to divide the magnetic field into
poloidal and toroidal components, ${\mathbf B} = {\mathbf B}_p + B_{\varphi} {\mathbf e}_{\varphi}$,
where the poloidal component can be written as
\begin{equation}
{\mathbf B}_p = \frac{1}{r} \nabla \Psi(r,z) \times {\mathbf e}_{\varphi}.
\label{eq:B_pPsi}
\end{equation}
We have adopted the cylindrical coordinate system $(r, \varphi, z)$. From Eq. (\ref{eq:B_pPsi}) we have
${\mathbf B} \cdot \nabla \Psi = 0$, i.e., $\Psi$ is constant along each magnetic field line.
$\Psi(r,z)$ is called {\it flux function}, representing the toroidal component of the 
vector potential times $r$ or the total magnetic flux penetrating within a circle of $r$ for a given $z$.

Eq. (\ref{eq:maxwell_E}) and Eq. (\ref{eq:mhd}) guarantee that $E_{\varphi} = 0$ and
${\mathbf E} \cdot {\mathbf B}=0$, respectively. Then the electric field can be written as
\begin{equation}
{\mathbf E} = -\frac{1}{c} \Omega \nabla \Psi.
\end{equation}
The toroidal component of Eq. (\ref{eq:maxwell_E}) leads to ${\mathbf B} \cdot \nabla \Omega = 0$, i.e.,
$\Omega = \Omega(\Psi)$ is constant along each magnetic field line. We also have a form
${\mathbf E} = (r\Omega {\mathbf B}_p/c)\times {\mathbf e}_{\varphi}$.

Dividing the particle velocity field into poloidal and toroidal components, 
${\mathbf v} = {\mathbf v}_p + v_{\varphi}{\mathbf e}_{\varphi}$, the ideal MHD condition gives us
\begin{eqnarray}
{\mathbf v}_p = \kappa {\mathbf B}_p  \label{eq:v_pB_p}, \label{eq:v_pB_p} \\
v_{\varphi} - \kappa B_{\varphi} = r \Omega(\Psi), \label{eq:Omega} 
\end{eqnarray}
where $\kappa$ is a function of $(r,z)$.

Eq. (\ref{eq:eoc}) together with Eqs. (\ref{eq:maxwell_B}) and (\ref{eq:v_pB_p}) reduce to
\begin{equation}
4 \pi \rho \Gamma \kappa = \eta(\Psi),
\label{eq:eta}
\end{equation}
where $\eta = \eta(\Psi)$ is constant along each magnetic field line and represents
the mass flux per unit magnetic flux.

Finally we divide Eq. (\ref{eq:eom}) into the direction parallel to ${\mathbf B}_p$, 
the azimuthal direction, and the direction normal to ${\mathbf B}_p$.
The component parallel to ${\mathbf B}_p$ is
\begin{equation}
u_p \frac{\partial}{\partial l} u_p = \frac{\sin\theta}{r} u_{\varphi}^2
-\frac{B_{\varphi}}{4\pi \rho c^2 r} \frac{\partial}{\partial l} (r B_{\varphi}),
\label{eq:eom_l}
\end{equation}
where $l$ is the coordinate of the direction of ${\mathbf B}_p$, and $\theta$ is the angle
between the $l$ and $z$ directions, i.e., $\sin\theta = \partial r/\partial l$.
The azimuthal component is 
\begin{equation}
u_p \frac{\partial}{\partial l} (r u_{\varphi}) 
= \frac{B_p}{4\pi \rho c^2} \frac{\partial}{\partial l}(r B_{\varphi}),
\end{equation}
which reduces to
\begin{equation}
r \Gamma v_{\varphi} - \frac{r B_{\varphi}}{\eta(\Psi)} = {\mathcal L}(\Psi),
\label{eq:L}
\end{equation}
where ${\mathcal L}(\Psi)$ is constant along each magnetic field line and represents
the total angular momentum flux per unit mass flux.
Differentiating Eq. (\ref{eq:L}) with $l$ and substituting it into Eq. (\ref{eq:eom_l})
together with Eq. (\ref{eq:Omega}), we obtain
\begin{equation}
\frac{\partial}{\partial l} \Gamma = \frac{\Omega}{c^2 \eta} \frac{\partial}{\partial l} (r B_{\varphi}),
\label{eq:diff_E}
\end{equation}
which means
\begin{equation}
\Gamma - \frac{r B_{\varphi} \Omega(\Psi)}{c^2 \eta(\Psi)} = {\mathcal E}(\Psi),
\label{eq:E}
\end{equation}
where ${\mathcal E}(\Psi)$ is constant along each magnetic field line and represents the 
total energy flux per unit rest energy flux. The ratio of Poynting to kinetic energy flux is given by
\begin{equation}
\sigma = \frac{-rB_{\varphi} \Omega}{c^2\eta \Gamma} = \frac{{\mathcal E} - \Gamma}{\Gamma}.
\label{eq:sigma}
\end{equation}
The component of the equation of motion normal to ${\mathbf B}_p$, i.e., transfield force balance equation,
is given in \cite{lyubarsky09}, which will not be treated in detail in this paper. 

\subsection{Bernoulli equation}

If $B_p(\Psi, r)$ is given, one can solve $\rho, \Gamma, \kappa, v_{\varphi},$ and $B_{\varphi}$
as functions of $(\Psi, r)$ with $\eta(\Psi), \Omega(\Psi), {\mathcal L}(\Psi),$ and ${\mathcal E}(\Psi)$
as parameters. Eqs. (\ref{eq:Omega}), (\ref{eq:eta}), (\ref{eq:L}), (\ref{eq:E}), and a relation
$\Gamma^2 - u_p^2 - u_{\varphi}^2 = 1$ reduce to an equation for $u_p$ 
as a function of $x \equiv r/r_{\rm lc}$ for a fixed $\Psi$,
\begin{equation}
u_p^2+1 = {\mathcal E}^2 \frac{x^2(1-x_A^2-M^2)^2-[x^2(1-x_A^2)-M^2 x_A^2]^2}{x^2(1-x^2-M^2)^2},
\label{eq:Bernoulli}
\end{equation}
which is called {\it Bernoulli equation} \cite{camenzind86}. Here $r_{\rm lc} \equiv c/\Omega$
is the radius of the light cylinder, 
\begin{equation}
x_A = \sqrt{\frac{{\mathcal L}\Omega}{{\mathcal E} c^2}}
\end{equation}
is the radius of the Alfv\'{e}n point divided by $r_{\rm lc}$, and 
\begin{equation}
M = \sqrt{\Gamma \kappa \eta}
\end{equation}
is the Alfv\'{e}n Mach number.
At the Alfv\'{e}n point, the denominator and numerator of Eq. (\ref{eq:Bernoulli}) both vanish, i.e., 
$1-x_A^2=M|_{x=x_A}^2$. Since $M|_{x=x_A}^2 > 0$, we have $x_A < 1$ in general.

Derivative of the Bernoulli equation with $x$ has a critical point where the poloidal 4-velocity
equals to fast magnetosonic speed,
\begin{equation}
u_p^2 = \frac{B_p^2(1-x^2) + B_{\varphi}^2}{4\pi \rho c^2} \equiv u_f^2.
\end{equation}
This is called the fast point \cite{camenzind86}. A solution
of $u_p$ growing towards infinity has to pass both Alfv\'{e}n and fast points. 
We call this a {\it wind solution}.
Since we have assumed that the flow is cold and neglected effects of gravitational field,
slow magnetosonic point does not appear.
Eq. (\ref{eq:Bernoulli}) can be rewritten as the fourth order algebraic equation 
for $u_p$, which can be solved for given $B_p(\Psi, x)/\eta(\Psi)$, $x_A (\Psi)$, and ${\mathcal E}(\Psi)$
\cite{camenzind86}. 
Corresponding 
$v_{\varphi}/c$ and $B_{\varphi}/(c\eta{\mathcal E})$ can be calculated by
\begin{eqnarray}
\frac{v_{\varphi}}{c} = \frac{x^2(1-x_A^2)-M^2x_A^2}{x(1-x_A^2-M^2)}, \\
\frac{B_{\varphi}}{c\eta{\mathcal E}} = \frac{x^2-x_A^2}{x(1-x^2-M^2)}.
\end{eqnarray}

For given $B_p (r)$, $\eta$, $\Omega$, and ${\mathcal L}$ along a field line, a unique value of
${\mathcal E}$ is determined for a wind solution. 
More physically, when $B_p(r)$, $\eta$, $\Omega$, and the velocity at the inlet $u_{p, {\rm in}}$ 
are given along a field line, Eqs. (\ref{eq:Omega}), (\ref{eq:L}), (\ref{eq:E}), and 
$\Gamma^2-u_p^2-u_{\varphi}^2=1$ evaluated at the inlet, together with the condition of the fast point,
determine the values of ${\mathcal L}$ and ${\mathcal E}$ and the solution of $u_p(r)$.
 
If the solutions for all the range of $\Psi$ satisfy the transfield force balance equation and
certain boundary conditions, they are a solution of the whole outflow structure, although we 
will not treat the whole structure in this paper.

\subsection{Asymptotic relations}

Asymptotic forms of various quantities beyond the light cylinder to the lowest order are easily
found along each magnetic field line \cite{lyubarsky09,vlahakis03}. 
Eliminating $B_{\varphi}$ from Eqs. (\ref{eq:L}) and (\ref{eq:E}), we obtain
\begin{equation}
\frac{v_{\varphi}}{c} = \frac{c}{r \Omega} \left(1- \frac{{\mathcal E}-{\mathcal L}\Omega/c^2}{\Gamma}
\right).
\end{equation}
We may assume that $v_{\varphi, {\rm in}} \ll c$ and $r_{\rm in} \Omega < c$ at the inlet, and then we obtain
${\mathcal E} - {\mathcal L}\Omega/c^2 \approx \Gamma_{\rm in}$. Thus we may write
\begin{equation}
\frac{v_{\varphi}}{c} = \frac{c}{r\Omega} \left(1- \frac{\Gamma_{\rm in}}{\Gamma}\right).
\label{eq:v_phi}
\end{equation}
For a wind solution in which $\Gamma \gg 1$ at $r \gg r_{\rm lc}$, we have $v_{\varphi} \propto r^{-1}$.

Since $v_p \approx c$ and $v_{\varphi} \ll c$ at $r \gg r_{\rm lc}$, we have from Eq. (\ref{eq:Omega})
\begin{equation}
\frac{-B_{\varphi}}{B_p} \approx \frac{r\Omega}{c} \gg 1.
\label{eq:B_phiB_p}
\end{equation}
In other words, the poloidal velocity dominates the total velocity, while the toroidal field dominates
the total magnetic field beyond the light cylinder. Eqs. (\ref{eq:E}) and (\ref{eq:B_phiB_p}) give us
\begin{equation}
{\mathcal E} \approx \Gamma + \frac{\Omega^2}{\eta c^3}B_p r^2,
\label{eq:approx_E}
\end{equation}
for $r \gg r_{\rm lc}$.
This equation reads that $\Gamma$ grows when $B_p r^2$ decreases along a field line
beyond the light cylinder. This can be also understood by Eq. (\ref{eq:eom_l}), in which the last
term of the right-hand side represents the Lorentz force. This force accelerates the fluids when
$|B_{\varphi}| r$ ($\propto B_p r^2$ at $r \gg r_{\rm lc}$) decreases along a field line.

For monopole magnetic field structure, i.e., $B_p r^2 = {\rm const}$, 
the flow cannot be efficiently accelerated. More specifically, in this case, the fast point is located
at infinity \cite{michel69,camenzind86}. 

For later calculations, we show two useful equations for $r \gg r_{\rm lc}$ here. 
Eliminating $v_p$ and $v_{\varphi}$
from the relation $(v_p/c)^2 + (v_{\varphi}/c)^2 + 1/\Gamma^2 = 1$ by using 
Eqs. (\ref{eq:Omega}) and (\ref{eq:v_phi}), we obtain 
\begin{equation}
\frac{B_{\varphi}^2}{B_p^2} = \frac{r^2\Omega^2}{c^2} 
\left[ 1+ \frac{1}{\Gamma^2} - \frac{c^2}{r^2\Omega^2} + {\mathcal O}\left(
\frac{1}{\Gamma^4}, \frac{c^2}{r^2\Omega^2\Gamma^2}, \frac{c^4}{r^4\Omega^4}\right)\right].
\label{eq:app_Bphi}
\end{equation}
We also have
\begin{equation}
\frac{v_p^2}{c^2} = 1-\frac{1}{\Gamma^2}-\frac{c^2}{r^2\Omega^2}+
\frac{2c^2}{r^2\Omega^2}\frac{\Gamma_{\rm in}}{\Gamma} + {\mathcal O}\left(
\frac{c^2}{r^2\Omega^2\Gamma^2}\right).
\label{eq:app_vp}
\end{equation}

\subsection{Wind solutions}
\label{subsec:wind}

\begin{figure}[t]
\centering\includegraphics[scale=0.8]{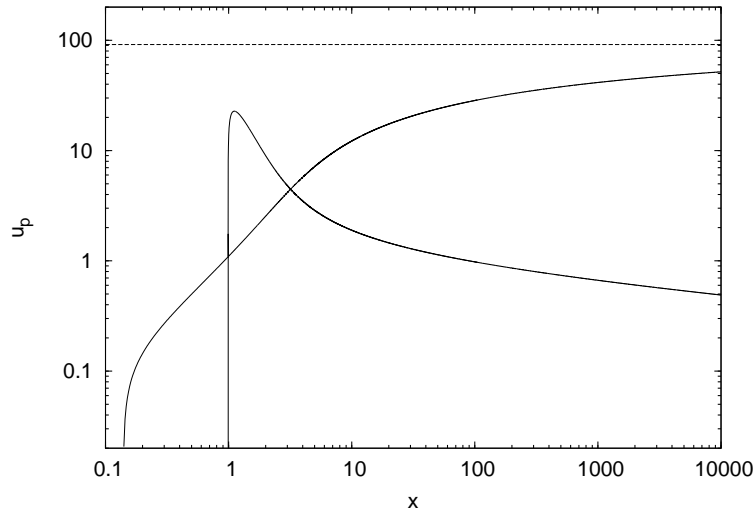}
\caption{
Solution of the Bernoulli equation with $B_p x^2 \propto x^{-q}$ with $q=0.1$. The parameters are
$B_p(x=1)/(c\eta)=100$, $x_A = 0.995$, and ${\mathcal E} = 91.6$. The dashed line
represents the maximum level of the Lorentz factor, $\Gamma \approx {\mathcal E}$.
}
\label{fig:fen_up}
\end{figure}

\begin{figure}[t]
\begin{minipage}{0.5\hsize}
\begin{center}
\includegraphics[scale=0.6]{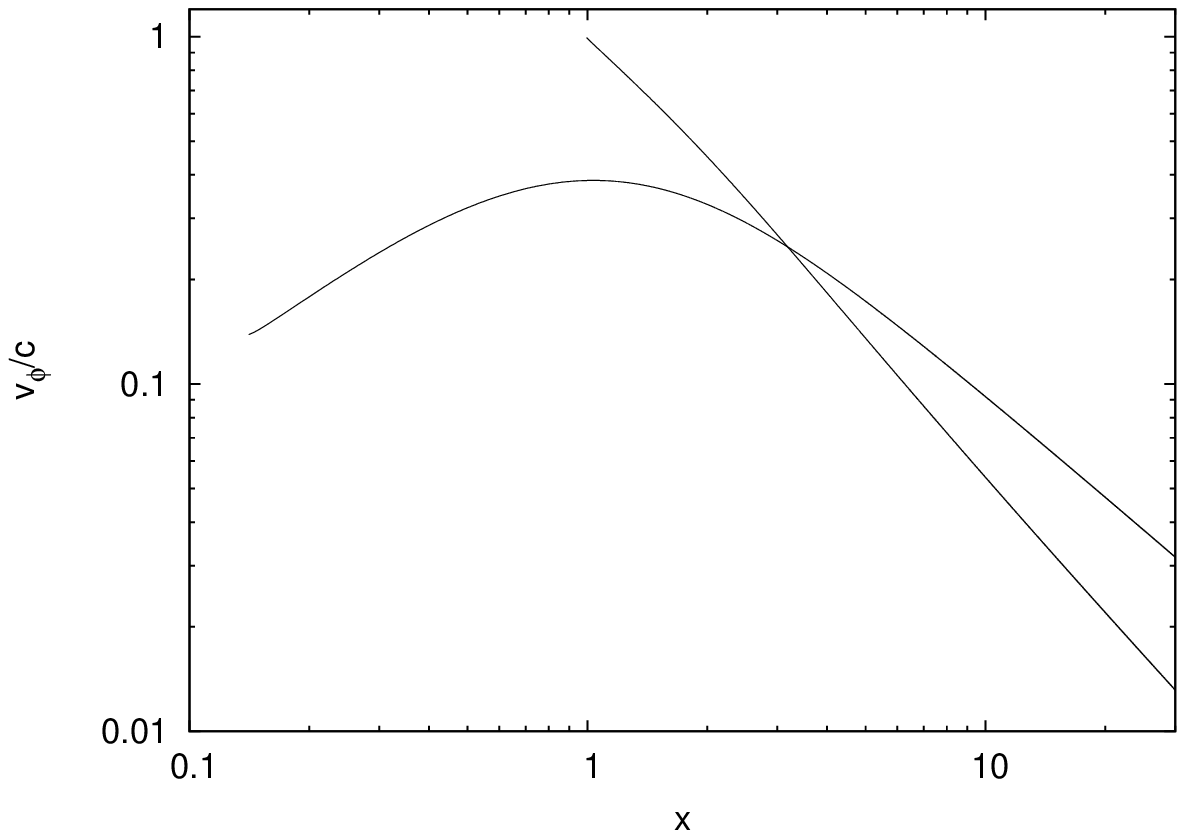}
\end{center}
\end{minipage}
\begin{minipage}{0.5\hsize}
\begin{center}
\includegraphics[scale=0.6]{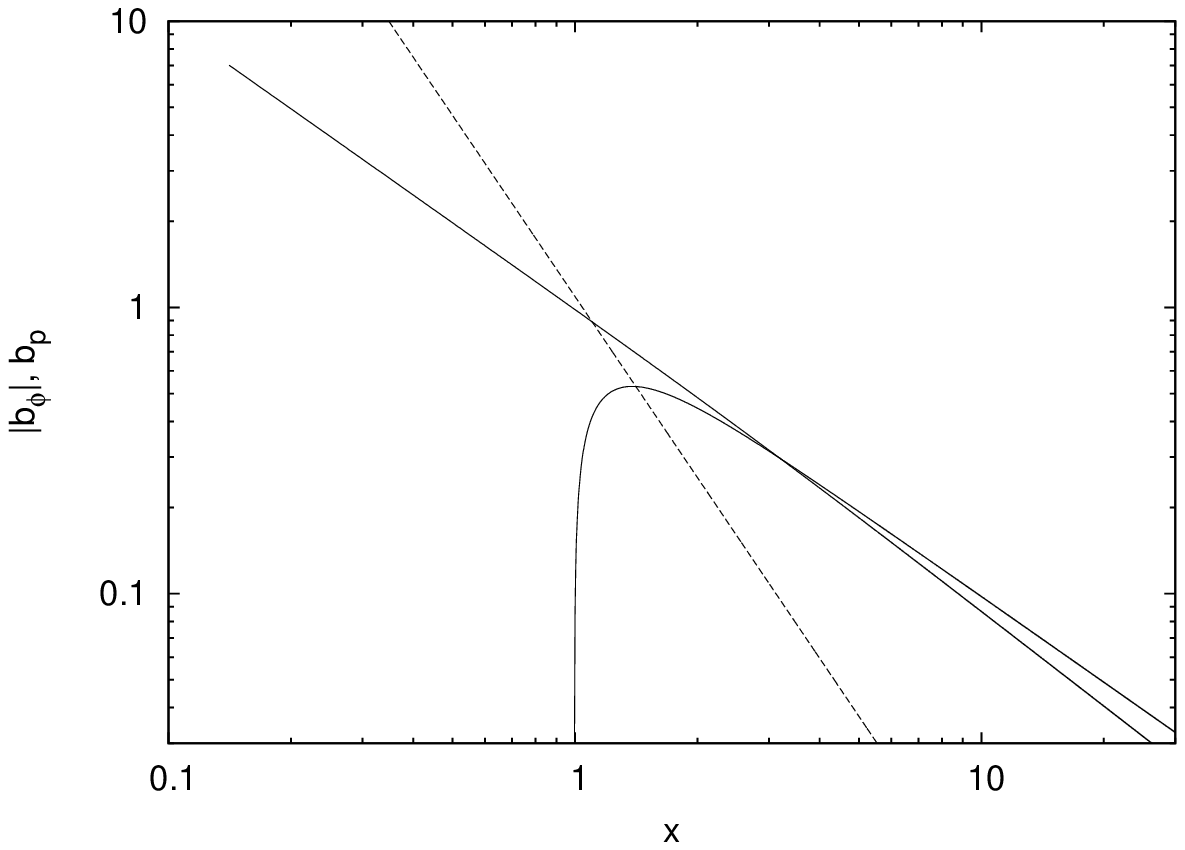}
\end{center}
\end{minipage}
\caption{
Solution for $v_{\varphi}/c$ (left) and $|b_{\varphi}| \equiv |B_{\varphi}|/(c\eta{\mathcal E})$ (right; solid line) 
and $b_p \equiv B_p/(c\eta{\mathcal E})$ (right; dashed line) corresponding to Figure \ref{fig:fen_up}.
}
\label{fig:fen_vphi}
\end{figure}

Fendt \& Ouyed (2004) \cite{fendt04} solved Eq. (\ref{eq:Bernoulli}) for 
cases of $B_p r^2 \propto r^{-q}$ with $q>0$. Following this formulation, we also solve 
Eq. (\ref{eq:Bernoulli}) and show a wind solution for $q=0.1$ in Figure \ref{fig:fen_up}.
The parameters are given as $B_p(x=1)/(c\eta)=100$, $x_A = 0.995$, and ${\mathcal E}=91.6$.
Note that if $v_{\varphi} \ll r\Omega$ and $\Gamma \ll {\mathcal E}$ at $r = r_{\rm lc}$ (i.e., $x=1$),
we have $x_A \sim 1$ and ${\mathcal E} \sim B_p(x=1)/c \eta$ from Eqs. (\ref{eq:L}) and (\ref{eq:E}).\footnote{
The values of $0.9945 < x_A < 1$ with the same values of the other 
parameters lead to wind solutions beginning at $0.1 < x < 1$.
If we set $x_A < 0.9945$, we obtain a wind solution beginning with a finite value of $u_p$. In other words,
$\Gamma \to {\mathcal E} (1-x_A^2)$ for $x \to 0$ as implied from Eqs. (\ref{eq:L}) and (\ref{eq:E}).
If $x_A$ is set to be larger than unity, there is no wind solution
passing the Alfv\'{e}n point from $x < 1$ to $x > 1$. 
See \cite{kennel83} for more details on the structure of the solutions of the Bernoulli equation. 
}
In reality, there are two solutions passing both Alfv\'{e}n and fast points, where $x=x_A$ and $x \simeq 3.2$,
respectively, as shown in Figure \ref{fig:fen_up}.
For one solution, the flow begins to accelerate at $x \ll 1$ and continue accelerating towards
infinity, while for the other solution, the flow begins at $x \approx 1$. We focus on the former
one, which we call a wind solution.
Our result is consistent with that in \cite{fendt04}.
As can be understood by Eq. (\ref{eq:approx_E}), the acceleration $d\Gamma/dx$ monotonically
decreases for cases of $B_p r^2 \propto r^{-q}$. This behavior is clearly seen in Figure \ref{fig:fen_up}.

Other quantities, $v_{\varphi}/c$ and $B_{\varphi}$, corresponding to Figure \ref{fig:fen_up} 
are shown in Figure \ref{fig:fen_vphi}. We can see that $v_{\varphi} \propto x^{-1}$ and 
$|B_{\varphi}|/B_p \approx x$ at $x \gg 1$, as discussed in the previous subsection.

For larger $q$, the energy conversion is more rapid, as demonstrated in \cite{fendt04} (which 
we also confirmed). Therefore for cases of $B_p r^2 \propto r^{-q}$, we can have an arbitrarily
efficient acceleration for large $q$.

\section{Magnetic nozzle}
\label{sec:nozzle}

In this section, we discuss behavior of the flow around the fast point a bit more deeply than
previous papers.
Komissarov et al. (2010) \cite{komissarov10} derived an asymptotic equation for $r \gg r_{\rm lc}$
from Eq. (\ref{eq:eom_l}), which is equivalent to
\begin{equation}
\left(\frac{1}{{\mathcal E}-\Gamma} - \frac{1}{\Gamma^3} \right) d\Gamma = \frac{-d{\mathcal S}}{\mathcal S},
\label{eq:komissarov_nozzle}
\end{equation}
where 
\begin{equation}
{\mathcal S} \equiv B_p r^2.
\end{equation}
This equation reads that in order to have $d\Gamma > 0$, ${\mathcal S}$ has to increase in the region of
$\Gamma < ({\mathcal E}-\Gamma)^{1/3}$ and vice versa, and $\Gamma = ({\mathcal E}-\Gamma)^{1/3}$
at the fast point. This behavior is quite similar to 
de Laval nozzle for a steady, one-dimensional, non-relativistic hydrodynamic flow \cite{landau59}, 
and often called {\it magnetic nozzle effect} \cite{begelman94}, where $1/{\mathcal S}$ corresponds to
the cross section of the magnetic nozzle. 
Some results of the numerical calculations by Komissarov et al. indeed show that
${\mathcal S}$ has a maximum (see Figure 10 of \cite{komissarov09}).
On the other hand, in the wind solution shown for the case of the monotonic decrease of ${\mathcal S}$, 
$B_p r^2 \propto r^{-q}$ with $q>0$, $u_p$ monotonically grows from non-relativistic to super-fast speed, 
passing through the fast point, as first demonstrated by \cite{fendt04} and also by Figure \ref{fig:fen_up}. 
This does not appear consistent with Eq. (\ref{eq:komissarov_nozzle}).

In order to solve this apparent inconsistency, we rewrite Eq. (\ref{eq:eom_l}) into a similar form as 
Eq. (\ref{eq:komissarov_nozzle}) but retaining the next order terms. As shown in Appendix, Eq. (\ref{eq:eom_l}) can 
be rewritten by using Eqs. (\ref{eq:Omega}), (\ref{eq:diff_E}), (\ref{eq:app_Bphi}), and (\ref{eq:app_vp}) as
\begin{equation}
\left(1-\frac{u_f^2}{u_p^2}\right) \frac{d\Gamma}{{\mathcal E}-\Gamma} + 
\frac{c^2 \Gamma_{\rm in}}{r^2 \Omega^2 \Gamma} \frac{d\Gamma}{u_p} +
\left(1+\frac{2\Gamma_{\rm in}}{\Gamma}\right)\frac{(-v_{\varphi} dv_{\varphi})}{v_p^2} 
= \frac{-d{\mathcal S}}{\mathcal S}.
\label{eq:our_nozzle}
\end{equation}
Note that this equation is valid only for $r > r_{\rm lc}$.
Clearly, this equation is not exactly the same as the case of de Laval nozzle. Even at and inside
the fast point (i.e., $u_p \leq u_f$), the second and third terms of the left-hand side allow $d\Gamma >0$
while $d{\mathcal S} <0$. Thus the monotonic decrease of ${\mathcal S}$ discussed in
\cite{fendt04} can have a solution of the flow passing through the fast point.
Eq. (\ref{eq:our_nozzle}) and its derivation indicate that the azimuthal direction of freedom allows
the solutions of the monotonic decrease of ${\mathcal S}$.
For $r \gg r_{\rm lc}$, however, the second and third terms of the left-hand side are negligible, and 
$u_f^2 \simeq B_{\varphi}^2/(4\pi \rho c^2 \Gamma^2) \simeq ({\mathcal E}-\Gamma)/\Gamma$,
and then we have Eq. (\ref{eq:komissarov_nozzle}). At the fast point, we have
\begin{equation}
\Gamma \sim ({\mathcal E}-\Gamma)^{1/3} ~~~ {\rm at} ~~~ r = r_f.
\label{eq:gamma_f}
\end{equation}
This property has been already shown in \cite{begelman94}.

\begin{figure}[t]
\begin{minipage}{0.5\hsize}
\begin{center}
\includegraphics[scale=0.6]{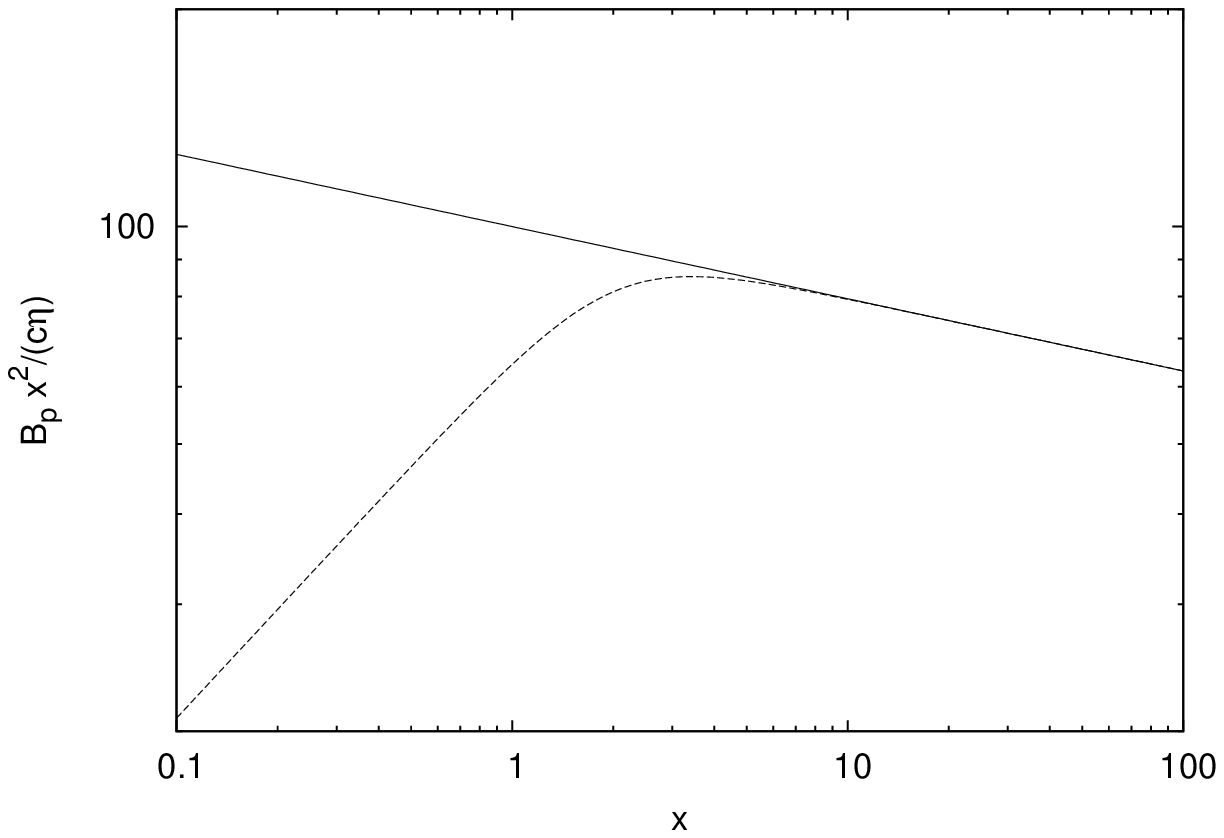}
\end{center}
\end{minipage}
\begin{minipage}{0.5\hsize}
\begin{center}
\includegraphics[scale=0.6]{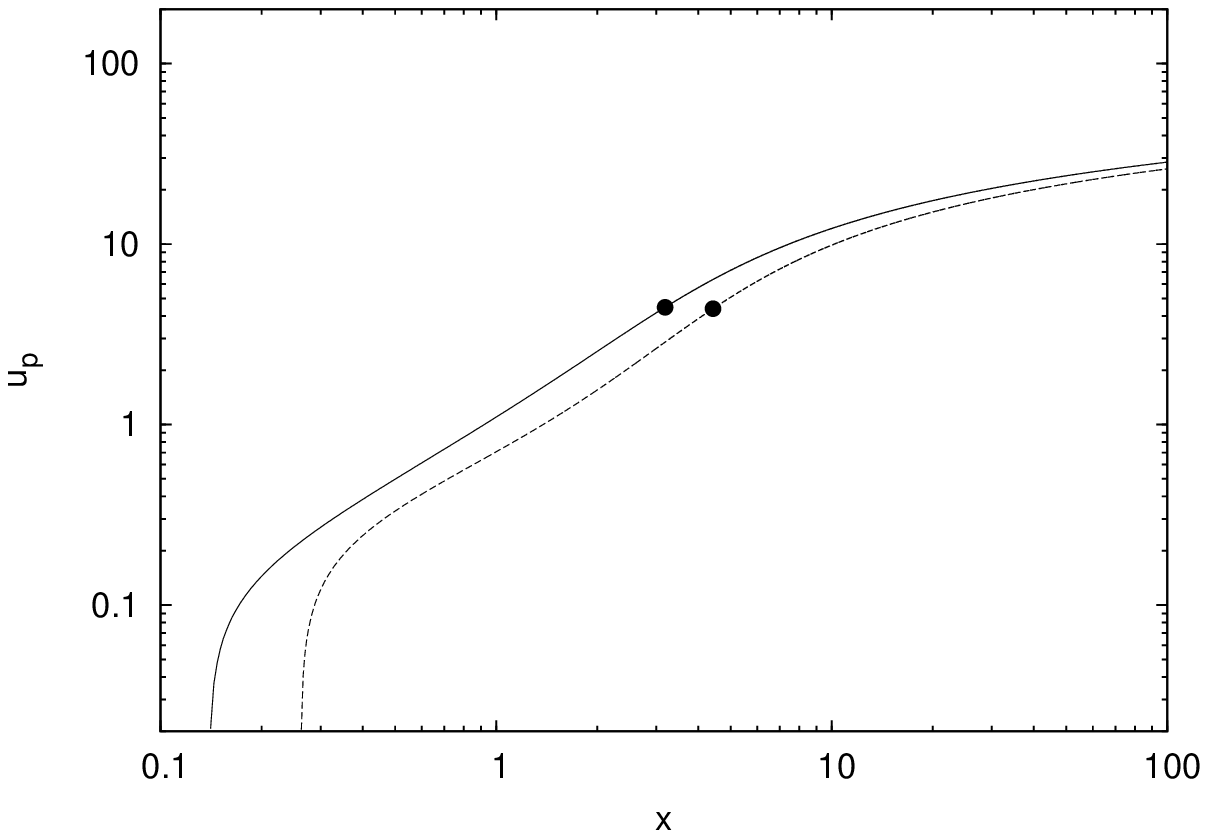}
\end{center}
\end{minipage}
\caption{
Different structures of ${\mathcal S}$ along a magnetic field line (left) and 
corresponding wind solutions for the Bernoulli equation (right). The solid line represents
the model of ${\mathcal S} \propto x^{-0.1}$ (same as Figure \ref{fig:fen_up})
and the dashed line represents that of Eq. (\ref{eq:broken}). 
The dots represent the fast points.
}
\label{fig:dpl}
\end{figure}

Now let us examine which shape of ${\mathcal S}$ leads to more efficient acceleration, the
monotonic decrease shape or the nozzle shape. 
In order to do this, we compare wind solutions of Eq. (\ref{eq:Bernoulli}) with 
different models of ${\mathcal S}$. The first model is ${\mathcal S} \propto x^{-q}$ with $q=0.1$ 
as shown in Figure \ref{fig:fen_up}, while the second one is a broken power-law form that has a 
maximum point,
\begin{equation}
{\mathcal S} \propto \left(\frac{x}{C} \right)^{p} \left[1+\left(\frac{x}{C}\right)^{s}\right]^{-(q+p)/s},
\label{eq:broken}
\end{equation}
with $p=0.5$, $q=0.1$, $s=3$, and $C=2.0$ (see Figure \ref{fig:dpl} left). For these parameters, 
${\mathcal S}$ has a maximum at $x \simeq 3.4$. We set $B_p/(c\eta)$ as taking the same value
as that of the first model at $x=30$, so that $B_p$ is almost the same for the two models far outside
the fast point. We set the same value of $x_A = 0.995$ for the two models.

Figure \ref{fig:dpl} (right) shows the two wind solutions. 
Here and hereafter we show the wind solutions only, not showing
the solutions of $u_p$ starting at $x \approx 1$ (see Figure \ref{fig:fen_up}), but instead
we indicate the fast points by dots. 
The first model, represented by the solid line
(which is the same as Figure \ref{fig:fen_up}), has ${\mathcal E} = 91.6$ while
the second model, represented by the dashed line, has ${\mathcal E} = 89.2$, a bit smaller than
the first model. The radius of the fast point is a bit larger in the second model. However,
the distances spent for the major energy conversions are very similar. This means that
the total acceleration efficiency hardly depends on the field structure inside the fast point.
In summary, one has $\Gamma \sim ({\mathcal E}-\Gamma)^{1/3}$ at the fast point in general,
and the total acceleration efficiency is determined by the poloidal field structure outside the fast point.

\section{Magnetic field shapes for efficient acceleration}
\label{sec:main}

In Section \ref{sec:basic}, we have seen that an arbitrarily efficient acceleration is available in the case
of $B_p r^2 \propto r^{-q}$ with $q>0$ along a field line. However, two-dimensional structure of
the poloidal field lines that gives such a scaling of $B_p(r)$ is not clear.
In this section, we show some types of the flux function $\Psi (r,z)$ for a rapid acceleration 
just beyond the fast point. 

Before doing it, we review an instructive example of the flux function that is given by
\begin{equation}
y + \zeta(\psi) = A(\psi)[x-\varpi(\psi)]^{a(\psi)},
\label{eq:lcl}
\end{equation}
where $y \equiv z/r_{\rm lc}$ and $\psi \equiv \Psi/\Psi_0$. This form of $\Psi$ has been discussed
in the literature \cite{chiueh91,vlahakis04}. 
$\zeta(\psi)$, $A(\psi)$, $\varpi(\psi)$, and $a(\psi)$ are arbitrary functions of $\psi$, but we
assume $\zeta' \geq 0$, $A' \leq 0$, $\varpi' \geq 0$, 
and $a' \leq 0$ to guarantee that $\Psi$ is smaller for more inside field lines, where the prime 
means the derivative with $\psi$. 
By calculating $B_r = (-1/r)\partial \Psi/\partial z$,
$B_z = (1/r)\partial \Psi/\partial r$, and $B_p = \sqrt{B_r^2 + B_z^2}$, we have
\begin{equation}
B_p x^2 = \frac{\Psi_0}{r_{\rm lc}^2} \left[\frac{\zeta'}{y+\zeta} - \frac{A'}{A} - a'\ln(x-\varpi)
+ \frac{a \varpi'}{x-\varpi} \right]^{-1} \frac{x}{x-\varpi} 
\sqrt{\left(\frac{x-\varpi}{y+\zeta}\right)^2 + a^2}.
\label{eq:lcl_bp}
\end{equation}
If $a>1$, the last factor has an effect of decreasing $B_p x^2$ along the field line for the range of 
$(x-\varpi)/(y+\zeta) > a$, i.e., for the range of $B_r > B_z$. 
In the second factor, $\zeta'/(y+\zeta)$ and $a\varpi'/(x-\varpi)$
increase $B_p x^2$, $A'/A$ does not change $B_p x^2$, and $-a'\ln(x-\varpi)$ decreases
$B_p x^2$. Thus only for $a'\neq 0$, a long-lasting decrease of
$B_p x^2$ is realized, although it decreases only logarithmically. 
Below we discuss a similar type of the flux function with $a' \neq 0$ as Type 1.

\subsection{Model and parameters}

\begin{figure}[t]
\centering\includegraphics[scale=0.6]{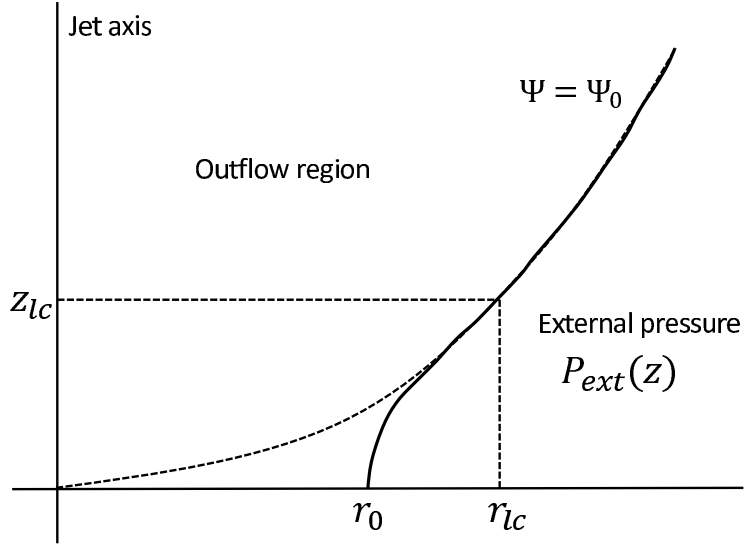}
\caption{
Schematic picture of our model for Type 1 and 2. The whole outflow region with 
non-zero poloidal magnetic flux $\Psi_0$ is confined by the external medium with thermal pressure 
and/or magnetic pressure of closed fields. We fix the shape of the external boundary as the solid line,
and discuss the flux function for the region just inside the boundary, 
$\Psi_0 - \delta\Psi < \Psi(r,z) \leq \Psi_0$, and wind solutions for the field line of $\Psi_0$.
The dashed curved line represents $y = A_0 x^{a_0}$ (except for Type 2-C),
where $y = z/r_{\rm lc}$ and $x = r/r_{\rm lc}$.
}
\label{fig:field_shape}
\end{figure}

Poloidal magnetic flux within a BH accretion system is not infinite 
in reality. We assume that the system has a finite region with non-zero poloidal magnetic flux around 
the axis, and this region drives a relativistic outflow. The finite magnetic flux is defined as $\Psi_0$
(see Figure \ref{fig:field_shape}).
The flow is confined by the pressure of the external medium, which includes thermal pressure 
and/or magnetic pressure of the closed fields. 
In this case the pressure balance condition
should be satisfied at the boundary \cite[cf.,][]{lyubarsky09}. Since the magnetic field measured
at the plasma rest frame is $B' = (B^2 - E^2)^{1/2}$, the boundary condition is given by
\begin{equation}
\left. \left(\frac{B_p^2 + B_{\varphi}^2 - E^2}{8\pi}\right)\right|_{\Psi = \Psi_0} = P_{\rm ext}(z).
\label{eq:boundary}
\end{equation}

We fix the shape of the external boundary, and discuss the flux function for the region 
just inside the external boundary, $\Psi_0 - \delta\Psi < \Psi(r,z) \leq \Psi_0$, and 
wind solutions for the field line of $\Psi_0$.

We assume that the boundary at $r \gtrsim r_{\rm lc}$ has a shape of
\begin{equation}
y = A_0 x^{a_0}
\label{eq:shape}
\end{equation}
(i.e., $\Psi(r,z)=\Psi_0$ is equivalent to Eq.(\ref{eq:shape})), where $A_0$ and $a_0$ are constants 
(except for Type 2-C below). At $r \lesssim r_{\rm lc}$
the boundary and the last field line are assumed to deviate from this shape and
be anchored to a thin disk at $r_0$ with Keplerian rotation, as illustrated by the solid line in Figure \ref{fig:field_shape}.
However, for simplicity, we use wind solutions of the Bernoulli equation along the field line shaping as $y=A_0 x^{a_0}$ 
for $0<x<\infty$ (i.e., the dashed curved line in Figure \ref{fig:field_shape}). This treatment is 
justified since the acceleration efficiency does not strongly depend
on the field structure inside the fast point, as discussed in Section \ref{sec:nozzle}.
For wind solutions we obtain, $P_{\rm ext}(z)$ can be deduced by Eqs. (\ref{eq:boundary}) and (\ref{eq:shape}). 
 
The parameter $\Omega$ is given by
\begin{equation}
\Omega(\Psi_0) = \sqrt{\frac{GM}{r_0^3}} = \sqrt{\frac{r_{\rm s} c^2}{2 r_0^3}}.
\end{equation}
Then the light cylinder radius is
\begin{equation}
r_{\rm lc}(\Psi_0) = \frac{c}{\Omega(\Psi_0)} = \sqrt{\frac{2r_0^3}{r_{\rm s}}} = \sqrt{2} R_0^{3/2} r_{\rm s},
~~~~~ \left(R_0 \equiv \frac{r_0}{r_{\rm s}}\right).
\label{eq:R_0}
\end{equation}
For the field line with $R_0 = 4$, as an example, we have $r_{\rm lc} \simeq 10\;r_{\rm s}$.

The wind solution $u_p(x)$ and corresponding external pressure profile $P_{\rm ext}(x)$ depend on
the parameters $A_0$ and $a_0$, but not on $R_0$, which 
just converts $(x,y)$ to $(r,z)$ by using Eq. (\ref{eq:R_0}).
To have rough constraints on $A_0$ and $a_0$, we may use a simple opening
angle estimate at the emission site as $r_e/z_e \sim 0.1$. Substituting this relation
into $y=A_0 x^{a_0}$ with $a_0=2$, we have $\sqrt{2} R_0^{3/2} \times 10^2 \sim A_0 (z_e/r_{\rm s})$.
Observations suggest that $z_e \sim 10^3 - 10^4\;r_{\rm s}$. For $z_e/r_{\rm s} = 10^3$,
we have $A_0 \sim 0.4, 1,$ and $3$ for $R_0=2, 4,$ and $8$, respectively. For $z_e/r_{\rm s} = 10^4$,
we have $A_0 \sim 0.04, 0.1,$ and $0.3$ for $R_0=2, 4,$ and $8$, respectively. 
For $a_0 =1$, we have $A_0 \sim 10$ for any value of $R_0$.
Below we will adopt $A_0 = 0.1$ and $A_0 = 1$ with $a_0 = 2$ and for $A_0 = 10$ with $a_0 =1$
as the fiducial values, for which we show calculation results.

\subsection{Type 1}

First let us consider a type for which the flux function for $\Psi_0 - \delta\Psi < \Psi \leq \Psi_0$
is given by
\begin{equation}
y = D \left(\frac{x}{d}\right)^{a(\psi)},
\label{eq:ff1}
\end{equation}
where $a' < 0$ and $a(\psi = 1) = a_0$. The parameters $D$ and $d$ are constants.
This type is suggested to make a long-lasting acceleration from the discussion below Eqs. (\ref{eq:lcl}) 
and (\ref{eq:lcl_bp}).
The constant $d$ should be $\ll 1$, since all the magnetic field lines intersect
and the field strength would diverge at $x=d$. For the field line of $\Psi = \Psi_0$, we have
\begin{equation}
B_p x^2 = \frac{\Psi_0}{r_{\rm lc}^2} \frac{1}{(-a'_0) \ln(x/d)} \sqrt{\left(\frac{x}{y}\right)^2 + a_0^2},
\label{eq:type1}
\end{equation}
where we have defined $a'_0 \equiv a'(\psi = 1)$. 

\begin{figure}[t]
\begin{minipage}{0.5\hsize}
\begin{center}
\includegraphics[scale=0.6]{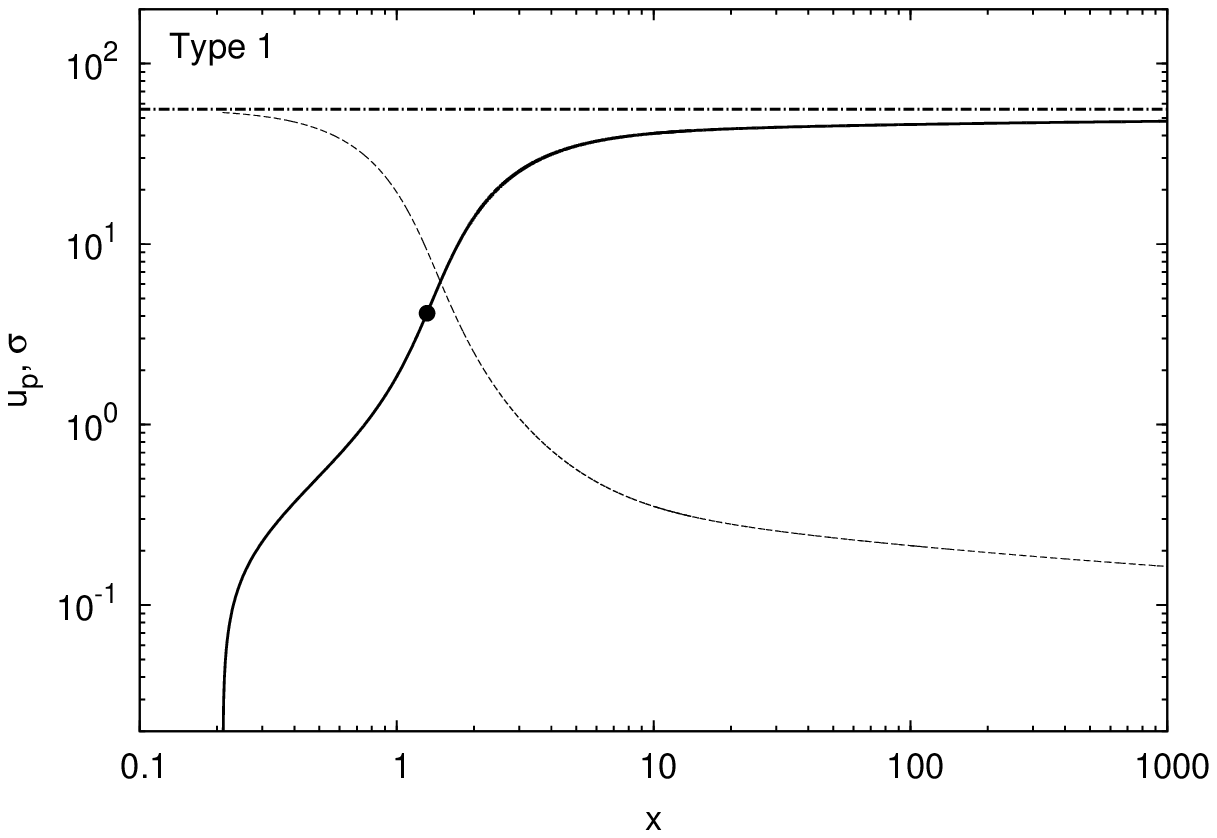}
\end{center}
\end{minipage}
\begin{minipage}{0.5\hsize}
\begin{center}
\includegraphics[scale=0.6]{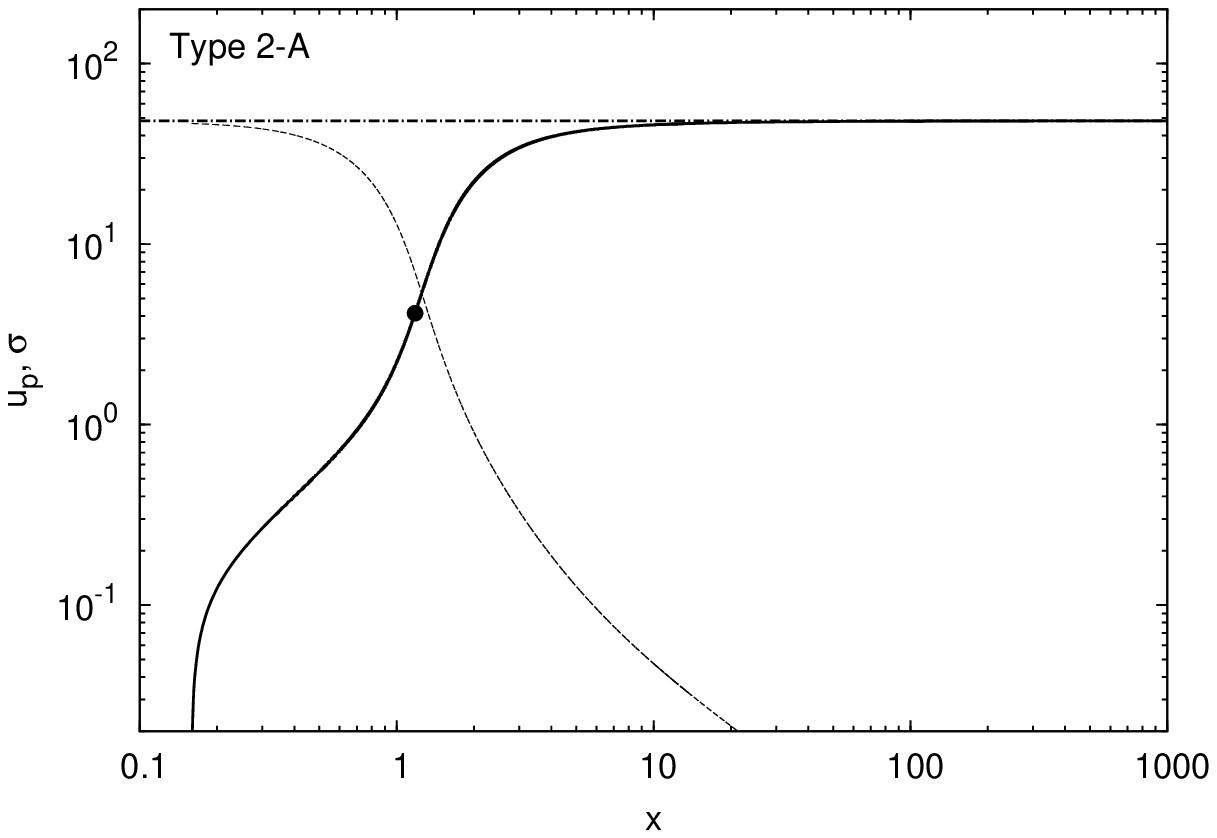}
\end{center}
\end{minipage}
\begin{minipage}{0.5\hsize}
\begin{center}
\includegraphics[scale=0.6]{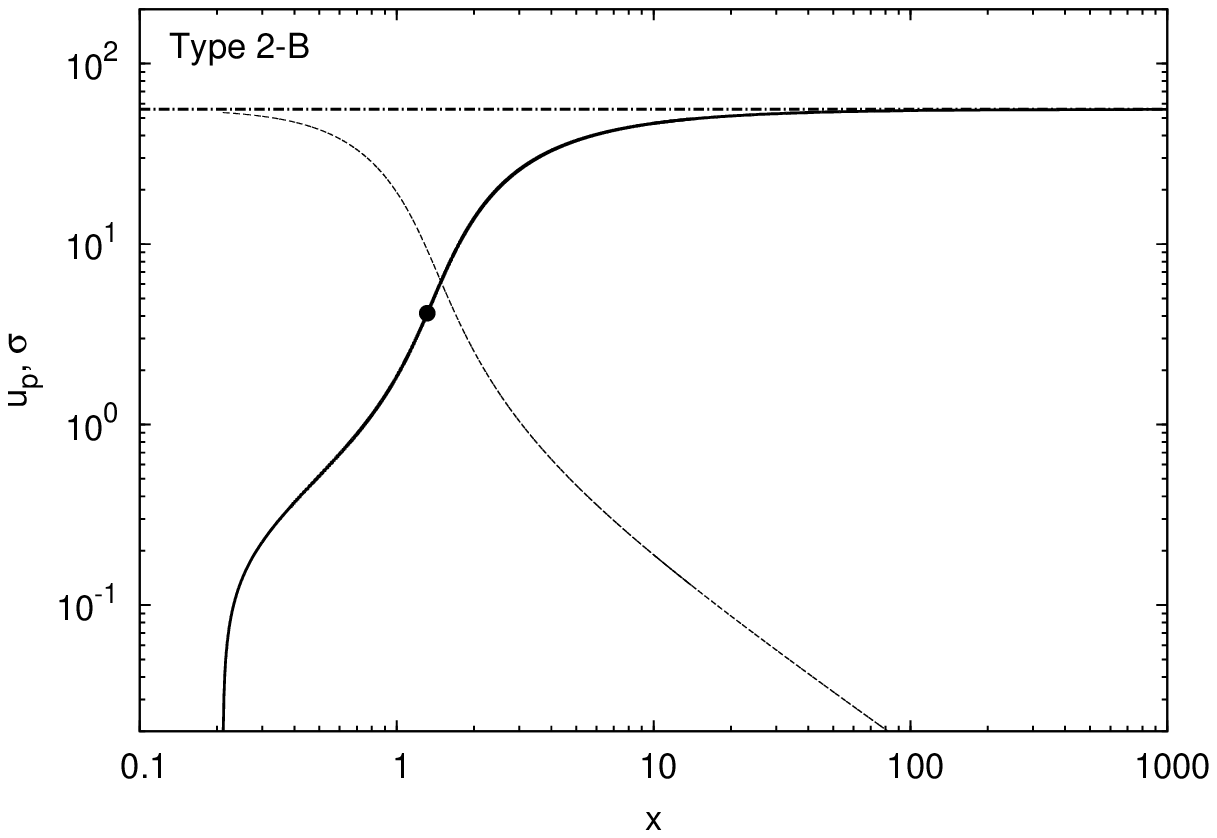}
\end{center}
\end{minipage}
\begin{minipage}{0.5\hsize}
\begin{center}
\includegraphics[scale=0.6]{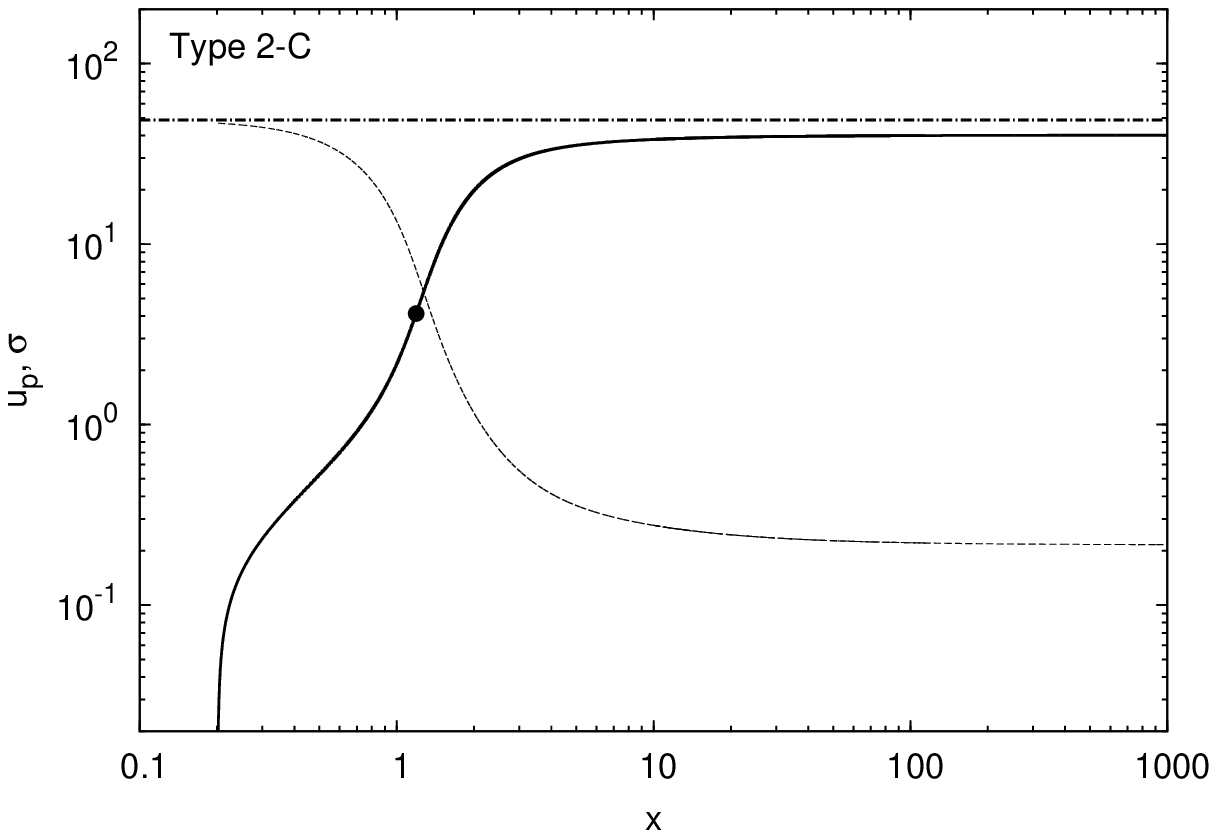}
\end{center}
\end{minipage}
\caption{
Wind solutions of the Bernoulli equation for $u_p$ (solid lines) and corresponding $\sigma$
as functions of $x$ (dashed line). The fast points are represented by dots, and the maximum level of the 
Lorentz factor, $\Gamma \approx {\mathcal E}$ by dot-dashed line.
Top left: $B_p x^2$ is given by Eq. (\ref{eq:type1}) and the parameters are
$B_p(x=1)/(c\eta)=100$, $d=10^{-2}$, $D=10^{-5}$, $a_0=2$, $x_A = 0.991$, and ${\mathcal E} = 55.8$. 
Top right: $B_p x^2$ is given by Eq. (\ref{eq:type2_2}) and the parameters are
$B_p(x=1)/(c\eta)=100$, $A_0=0.1$, $a_0=2$, $F_0=0$, $b=3$, $x_A = 0.990$, and ${\mathcal E} = 48.2$. 
Bottom left: $B_p x^2$ is given by Eq. (\ref{eq:type2_2}) and the parameters are
$B_p(x=1)/(c\eta)=100$, $A_0=1.0$, $a_0=2$, $F_0=0$, $b=3$, $x_A = 0.991$, and ${\mathcal E} = 55.8$. 
Bottom right: $B_p x^2$ is given by Eq. (\ref{eq:type2_1}) and the parameters are
$B_p(x=1)/(c\eta)=100$, $A_0=1.0$, $a_0=2$, $F_0=0.03$, $b=3$, $x_A = 0.990$, and ${\mathcal E} = 48.7$. 
}
\label{fig:type1_up}
\end{figure}

Figure \ref{fig:type1_up} (top left) shows a wind solution for $B_p x^2$ given by Eq. (\ref{eq:type1})
and a corresponding profile of $\sigma$ (see Eq. \ref{eq:sigma}) by the solid and dashed lines, respectively.
The parameters are given as $B_p(x=1)/(c\eta)=100$, $d=10^{-2}$, $D=10^{-5}$, $a_0=2$, $x_A = 0.991$, 
and ${\mathcal E} = 55.8$. This parameter choice corresponds to the field line shape of $\Psi = \Psi_0$
as Eq. (\ref{eq:shape}) with $a_0=2$ and $A_0=0.1$.
For $x/y > a_0$, i.e., $x<5$, $B_p x^2$ rapidly decreases, roughly scaling as $\propto x/y \propto x^{-1}$,
i.e., $B_p x^2 \propto x^{-q}$ with $q=1$, so that the acceleration is very rapid. This causes the fast 
point, where $\Gamma \approx {\mathcal E}^{1/3} \simeq 3.8$, to be closer to the Alfv\'{e}n point than 
the case of Figure \ref{fig:fen_up} ($q=0.1$).
For $x > 5$, however, the last factor of Eq. (\ref{eq:type1})
is roughly constant, so that $B_p x^2$ decreases only logarithmically. As a result, 
we have $\sigma \leq 0.2$ at $x \geq 170$, corresponding to $y \geq 2.9 \times 10^3$ and then
$z \geq 4 \times 10^3\;R_0^{3/2} r_{\rm s}$, although $\sigma \leq 0.1$ is realized only at $x > 10^4$, 
corresponding to $z > \sqrt{2} \times 10^7 R_0^{3/2} r_{\rm s}$, which is very large distance compared 
with the observational suggestions.

\begin{figure}[t]
\begin{minipage}{0.5\hsize}
\begin{center}
\includegraphics[scale=0.6]{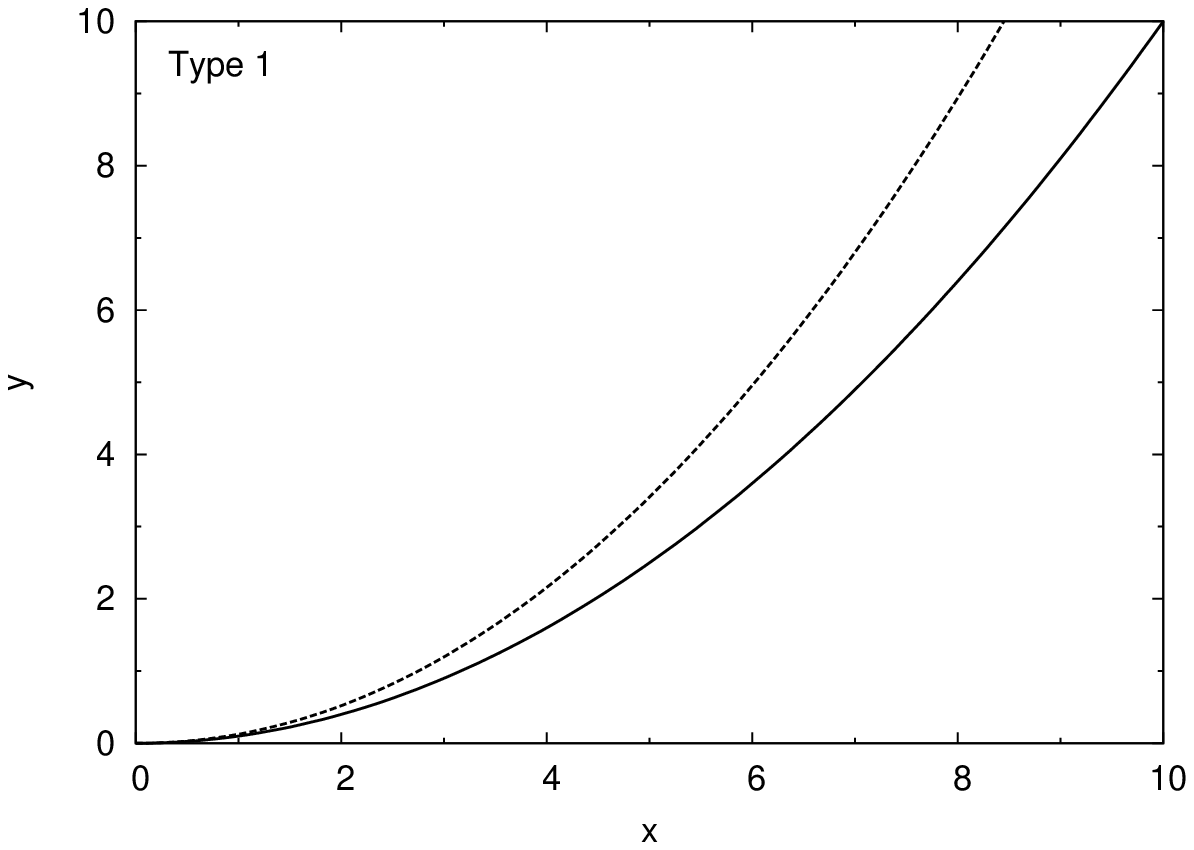}
\end{center}
\end{minipage}
\begin{minipage}{0.5\hsize}
\begin{center}
\includegraphics[scale=0.6]{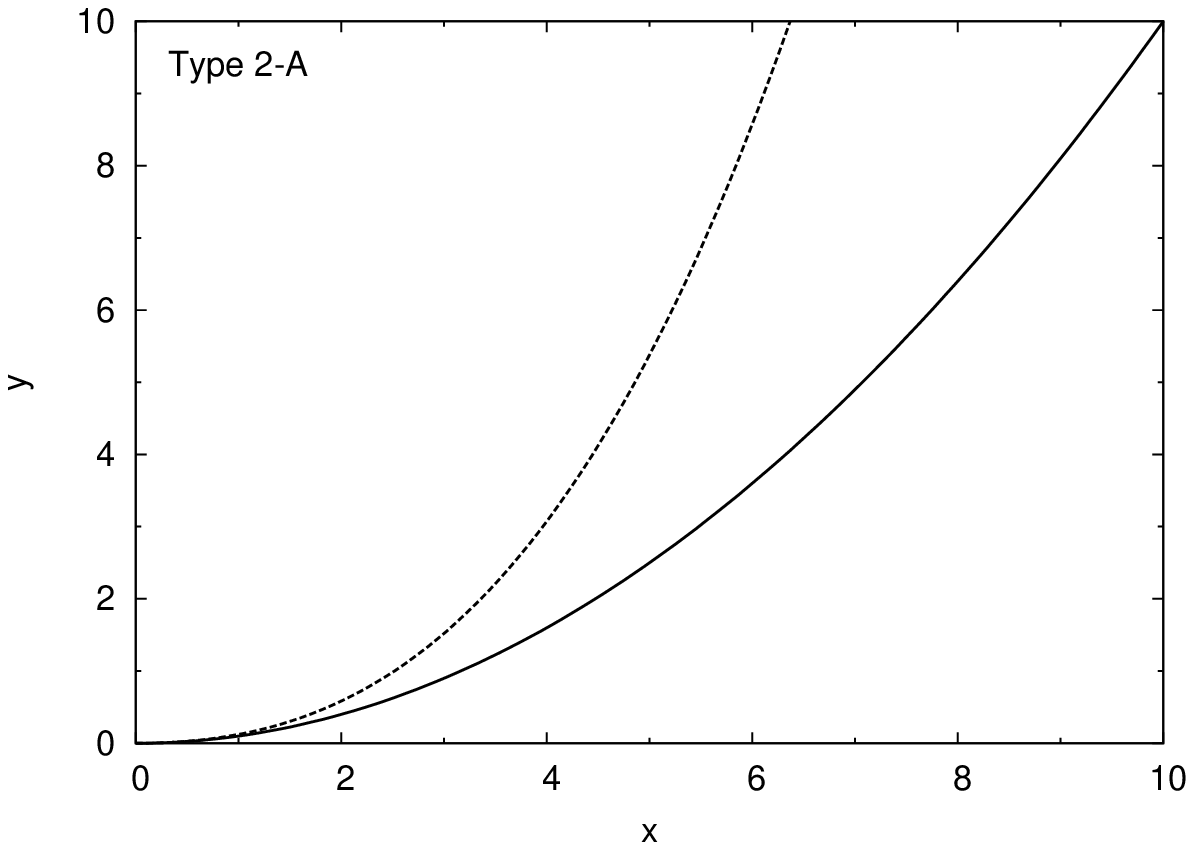}
\end{center}
\end{minipage}
\begin{minipage}{0.5\hsize}
\begin{center}
\includegraphics[scale=0.6]{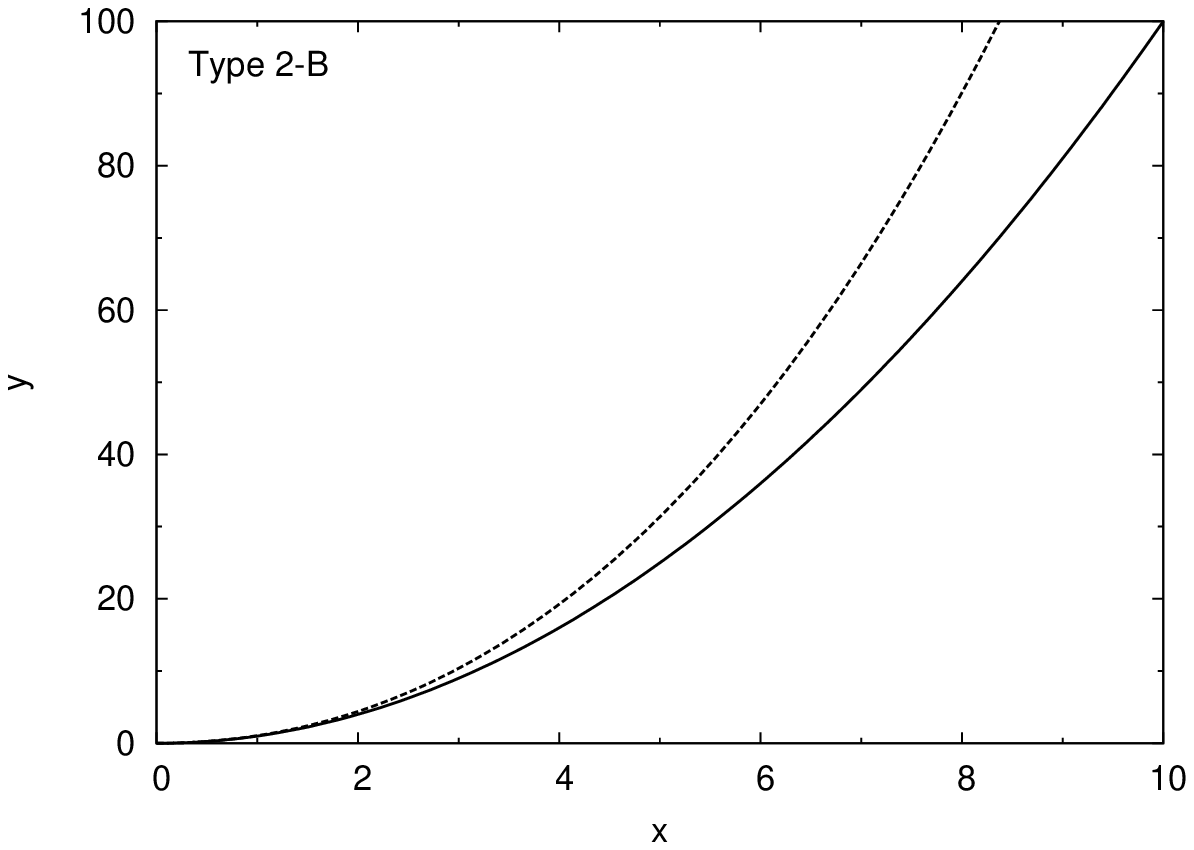}
\end{center}
\end{minipage}
\begin{minipage}{0.5\hsize}
\begin{center}
\includegraphics[scale=0.6]{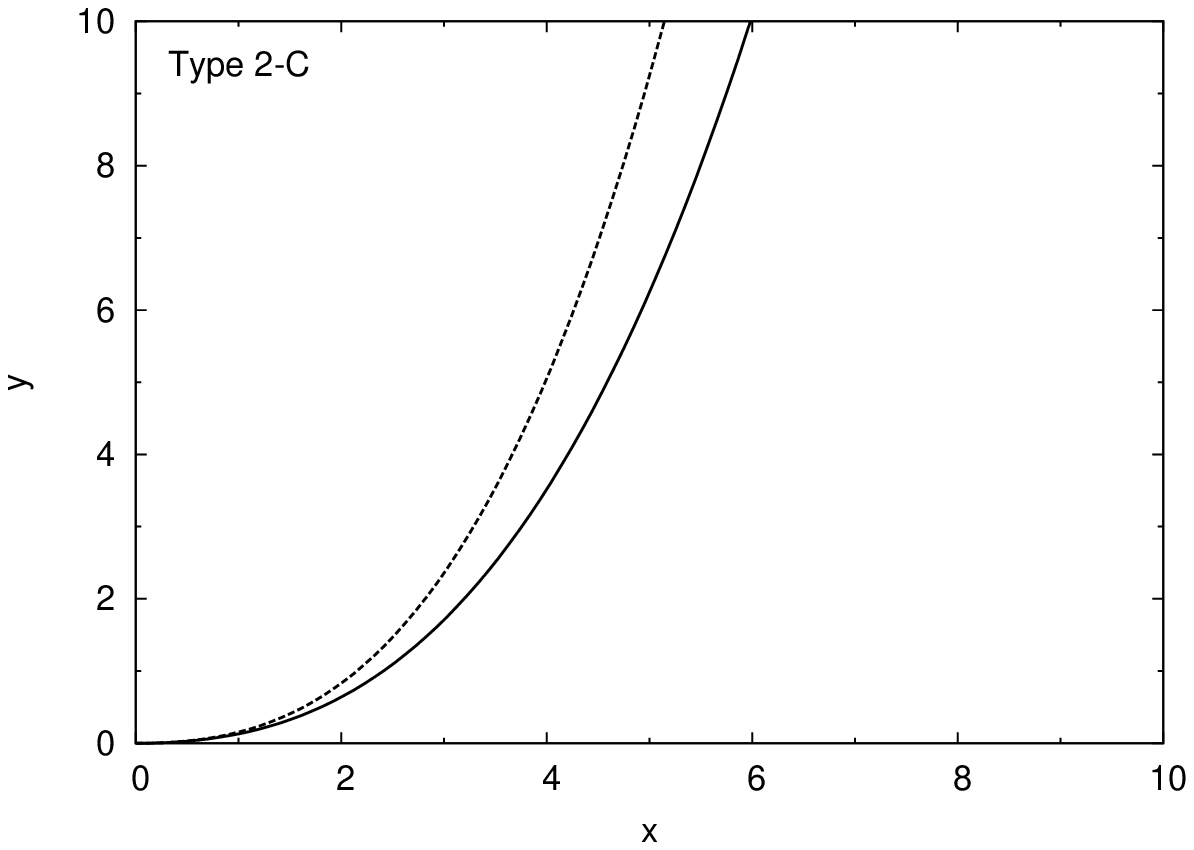}
\end{center}
\end{minipage}
\begin{minipage}{0.5\hsize}
\begin{center}
\includegraphics[scale=0.6]{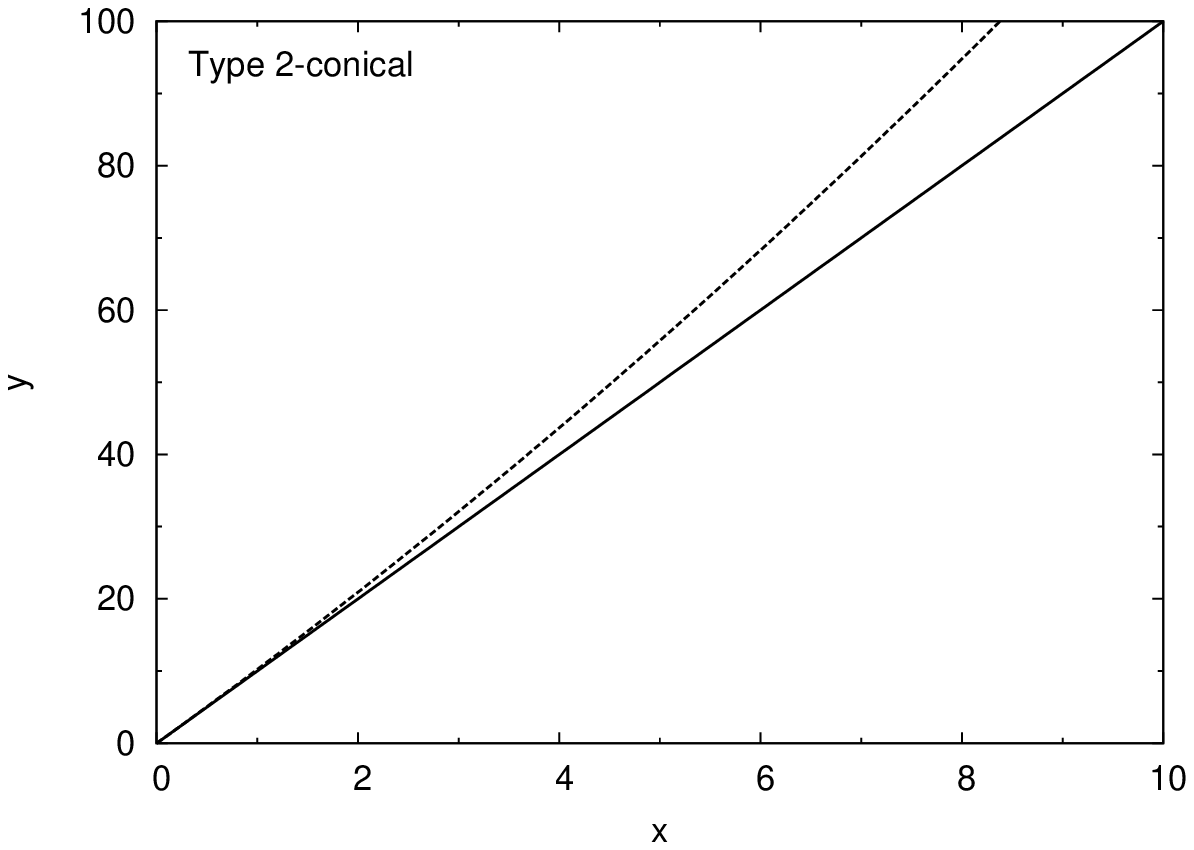}
\end{center}
\end{minipage}
\caption{
Poloidal field lines of $\psi = 0.95$ for the flux functions given by Eqs. (\ref{eq:ff1}) and 
(\ref{eq:ff2}) for which $\Psi_0/(c \eta r_{\rm lc}^2) \simeq 46$ is set (dashed lines). 
The top left, top right, middle left, middle right, and bottom panels 
correspond to Type 1, 2-A, 2-B, 2-C, and 2-conical, respectively.
The solid lines represent the external boundary $\psi=1$. Note that the vertical scales of 
the middle left and bottom panels are different from those of the other panels.
}
\label{fig:para}
\end{figure}

To illustrate the magnetic field structure of this type, we plot the field line of $\psi=0.95$
with the dashed line in Figure \ref{fig:para} (top left). This is the line of Eq. (\ref{eq:ff1})
with the power law index set to be $a(\psi) = a_0 + a'_0(\psi - 1)$. We have assumed
that $\Psi_0/(c\eta r_{\rm lc}^2) = 10 \times \ln(10^2) \simeq 46$ and given $a'_0$ by using 
Eq. (\ref{eq:type1}) evaluated at $x=1$ with $B_p(x=1)/(c\eta)=100$.

\begin{figure}[t]
\centering\includegraphics[scale=0.8]{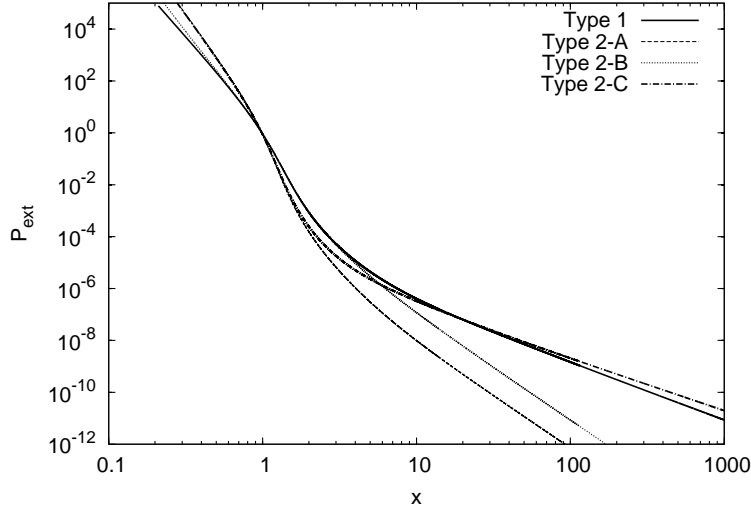}
\caption{
External pressures as functions of $x$ corresponding to the wind solutions in Figure \ref{fig:type1_up}.
The solid, dashed, dotted, and dot-dashed lines correspond to Type 1, 2-A, 2-B, and 2-C, respectively.
}
\label{fig:type1_P}
\end{figure}

Figure \ref{fig:type1_P} (solid line) shows the external pressure calculated by Eq. (\ref{eq:boundary}).
$P_{\rm ext}(x)$ decreases very steeply just beyond the light cylinder, and it roughly scales as $\propto x^{-2}$
far outside the fast point. This behavior can be understood by an approximate equation of Eq. (\ref{eq:boundary}),
\begin{equation}
P_{\rm ext} \approx \frac{B_p^2}{8\pi} \frac{r^2 \Omega^2}{c^2} \frac{1}{\Gamma^2},
\label{eq:approx_P}
\end{equation}
which is derived by using Eq. (\ref{eq:app_Bphi}) for $x>1$.
For the rapid acceleration region, $1<x<3$, the external pressure 
$P_{\rm ext} \propto B_p^2 r^2/\Gamma^2 \propto x^{-4}/\Gamma^2$
decreases by a factor of $\sim 10^{-5}$. For $x \gg 5$, $\Gamma \sim$ const, and the external 
pressure scales as $P_{\rm ext} \propto B_p^2 r^2 \propto x^{-2}/(\ln x)^2$.
Since $z \propto x^2$, we have $P_{\rm ext} \propto z^{-1}/(\ln z)^2$.

\subsection{Type 2}

Next we consider a flux function given by
\begin{equation}
y = A_0 x^{a_0} + F(\psi) x^b,
\label{eq:ff2}
\end{equation}
for $\Psi_0 - \delta \Psi < \Psi \leq \Psi_0$, where $a_0<b$, $F(\psi) < A_0$, 
$F'<0$, and $b$ is constant. For this flux function we have more efficient and rapid acceleration. 
Eq. (\ref{eq:ff2}) provides
\begin{equation}
B_p x^2 = \frac{\Psi_0}{r_{\rm lc}^2} \frac{A_0 x^{a_0-b}+F}{(-F')} \sqrt{\left(\frac{x}{y}\right)^2 + 
\left(\frac{A_0 a_0 x^{a_0-1}+Fbx^{b-1}}{A_0 x^{a_0-1}+Fx^{b-1}}\right)^2}.
\label{eq:type2_1}
\end{equation}
Choosing the boundary shape as Eq. (\ref{eq:shape}), we set $F_0 \equiv F(\psi = 1) = 0$ and obtain
for the field line of $\Psi_0$
\begin{equation}
B_p x^2 = \frac{\Psi_0}{r_{\rm lc}^2} \frac{A_0 x^{a_0-b}}{(-F_0')} \sqrt{\left(\frac{x}{y}\right)^2 + a_0^2},
\label{eq:type2_2}
\end{equation}
where we have defined $F'_0 \equiv F'(\psi = 1)$.

Figure \ref{fig:type1_up} (top right) shows a wind solution for $B_p x^2$ given by Eq. (\ref{eq:type2_2}).
The parameters are $B_p(x=1)/(c\eta)=100$, $A_0=0.1$, $a_0=2$, $b=3$, $x_A = 0.990$, and ${\mathcal E} = 48.2$,
and we call this case `Type 2-A'.
In this case, $B_p x^2$ decreases as a power law function of $x$, not logarithmically 
(i.e., $B_p x^2 \propto x^{-q}$ with $q = b-a_0 = 1$), even
outside the rapid decrease phase of $x/y > a_0$. As a result, we have $\sigma \leq 0.1$ as
early as $x \geq 6$. This corresponds to $z \geq 3.6 \times \sqrt{2} R_0^{3/2}\;r_{\rm s} 
\sim 40\;r_{\rm s}$ for $R_0 =4$, which is small enough to obtain $\sigma < 0.1$ at the
emission site, as suggested by observations. 

The dashed line in Figure \ref{fig:para} (top right) represents the poloidal field line of $\psi = 0.95$
for this type of the flux function. This line is plotted as Eq. (\ref{eq:ff2}) with 
$F(\psi) = F_0 + F'_0(\psi-1)$ and with the same value of $\Psi_0/(c\eta r_{\rm lc}^2)$ 
as that for Type 1, i.e., the dashed line in Figure \ref{fig:para} (top left). 
This clearly shows that the field lines of Type 2 expand sideways (and $B_p x^2$ decreases) 
more rapidly than those of Type 1.
 
Figure \ref{fig:type1_P} (dashed line) shows the corresponding structure of the
external pressure. It drops very rapidly at $1 < x < 3$ and decays as a power law of $x$ at $x \gg 5$,
similarly as Case 1. At $x \gg 5$, Eq. (\ref{eq:approx_P}) implies that $P_{\rm ext} \propto B_p^2 r^2 \propto x^{-4}
\propto z^{-2}$.

We also obtain a wind solution for the case of $A_0=1.0$ and show the result in Figure \ref{fig:type1_up}
(bottom left). 
We call this case `Type 2-B'.
For this parameter value, the rapid acceleration phase $x/y > a_0$ (i.e., $x < 0.5$) is so short that
the acceleration is slower than the case of Figure \ref{fig:type1_up} (top right). 
This means that more-collimated jets
have a slower fluid acceleration in this type of the flux function. Even though the last term of 
Eq. (\ref{eq:type2_2}) does not strongly contribute to decreasing $B_p x^2$, $u_p$ increases rapidly
just beyond $x \sim 1$. This is simply a generic property for the cases of $B_p x^2$ decreasing as 
a power law function of $x$, as discussed in Section \ref{subsec:wind}. While the acceleration is 
rapid in the initial phase, the acceleration 
$d\Gamma/dx$ decreases monotonically, as can be expected by Eq. (\ref{eq:approx_E}).
The range where $\sigma \leq 0.1$
is $x \geq 18$, which corresponds to $z \geq 3.2 \times 10^2 \times \sqrt{2} R_0^{3/2} r_{\rm s} \sim 
4 \times 10^3\;r_{\rm s}$ for $R_0 = 4$. This is comparable to the observational suggestion
$z_e \sim 10^3\;r_{\rm s}$.
For reference, we plot the the field line of $\psi = 0.95$ in Figure \ref{fig:para} (middle left)
and the external pressure profile in Figure \ref{fig:type1_P} (dotted line).

Next let us examine the cases for $a_0 = 1$, $b=2$, and $F_0 = 0$. 
The external boundary in this case has the conical shape. The field line of $\psi=0.95$
for $B_p(x=1)/(c\eta)=100$, $A_0 =10$, and
$\Psi_0/(c \eta r_{\rm lc}^2) \simeq 46$ is shown in Figure \ref{fig:type1_P} (bottom).
The wind solution we obtain for $x_A = 0.992$ and ${\mathcal E} = 58.2$ (not explicitly shown)
is quite similar to Figure \ref{fig:type1_up} (bottom left). 
This result is expected, because the last factor of Eq. (\ref{eq:type2_2})
is constant, not contributing to decreasing $B_p x^2$, and the power law index of $B_p x^2$ for $x$,
$-q = a_0-b = -1$ is the same as the case of Figure \ref{fig:type1_up} (bottom left). 
The range where $\sigma \leq 0.1$, $x \geq 18$, corresponds to $z \geq 1.8 \times 10^2 \times
\sqrt{2} R_0^{3/2} r_{\rm s} \sim 2 \times 10^3\;r_{\rm s}$, which is small enough to be 
consistent with the observational suggestions.

Finally, we consider the case in which $F_0 \neq 0$. In this case, the boundary
does not have the shape of Eq. (\ref{eq:shape}) but obeys Eq. (\ref{eq:ff2}) with $F_0 \neq 0$.
For instance, we adopt
the parameters as $A_0=0.1$, $a_0=2$, $F_0=0.03$, and $b=3$. We find a wind solution for $B_p x^2$
given by Eq. (\ref{eq:type2_1}) with
$B_p(x=1)/(c\eta)=100$, $x_A=0.990$, and ${\mathcal E} = 48.7$, and the result is shown 
in Figure \ref{fig:type1_up} (bottom). We call this case `Type 2-C'.
In this case, decrease of $B_p x^2$ stops at $x \sim A_0/F_0 \sim 3$, so that
the energy conversion saturates at $\sigma \simeq 0.22$. The external pressure far beyond the 
fast point obeys $P_{\rm ext} \propto B_p^2 r^2 \propto x^{-2}$ (see the dot-dashed line in 
Figure \ref{fig:type1_P}). 
For reference, we plot the field line of $\psi = 0.95$ in Figure \ref{fig:para} (middle right).

\section{Conclusion and Discussion}

Recently, numerical and analytical studies of global structures of steady, axisymmetric, 
relativistic MHD flows with specific external boundary conditions have been significantly 
developed \cite{komissarov07,komissarov09,tchekhovskoy09,lyubarsky09,lyubarsky10}.
They all showed that the energy conversion from Poynting into particle kinetic energy flux
is rapid up to the equipartition level, i.e., until $\sigma \sim 1$ is attained, but further conversion 
up to $\sigma \ll 1$ is very slow. As a result, the radius where $\sigma \lesssim 0.1$ is too
large, compared with observational suggestions that $\sigma \lesssim 0.1$ at the
emission site $z \sim 10^3 - 10^4\;r_{\rm s}$ \cite{inoue96,kino02,sikora05}.

In this context, we consider that it is important to clarify whether there can be 
magnetic field structures and boundary conditions for which rapid energy
conversion up to $\sigma \lesssim 0.1$ occurs. 
We have fixed the shape of the external boundary of the outflow (see Figure \ref{fig:field_shape}) 
and looked for general types of the flux function for the region just inside the boundary leading 
to $\sigma \lesssim 0.1$ soon after the fast point (without discussing the whole transverse 
structure of the outflow). Although we have solved the equation
of motion along the field line shaping as $z \propto r^{a_0}$ for $0 < r < \infty$ (dashed line
in Figure \ref{fig:field_shape}), those solutions are roughly applicable for the more realistic field line 
anchored at $r_0 < r_{\rm lc}$ (the solid line in Figure \ref{fig:field_shape}), because the 
acceleration efficiency is determined by the field structure outside the fast point,
as discussed in Section \ref{sec:nozzle} (see Figure \ref{fig:dpl}).
We have found a type of the flux function, given by Eq. (\ref{eq:ff2}) near the boundary, which 
describes the magnetic flux tubes expanding sideways rapidly as shown by 
the dashed lines in Figure \ref{fig:para} (top right) and (middle left),
and leads to a rapid 
energy conversion just beyond the fast point (see Figures \ref{fig:type1_up} (top right) and (bottom left)).
In this case we can have $\sigma \lesssim 0.1$ at $r \lesssim 10^3 - 10^4\;r_{\rm s}$, 
consistent with observational suggestions. For the flux function given by Eq. (\ref{eq:ff1}), 
which was often discussed in the literature \cite{chiueh91,vlahakis04}, the flux tubes
expand slower than the above function, 
as shown by the dashed line in Figure \ref{fig:para} (top left),
and the acceleration is much slower.

The poloidal velocity of the fluid is parallel to the poloidal magnetic field line, and the fluid
is accelerated by the Lorentz force 
when $|B_{\varphi}| r$ decreases along a field line (see Eq. \ref{eq:eom_l}). 
Generally, in the relativistic fluid, Poynting flux is not 
significantly converted into particle kinetic energy flux at the fast point 
(i.e., $\Gamma \approx {\mathcal E}^{1/3}$), so that the major energy conversion has to 
be realized beyond the fast point. The major conversion up to $\Gamma \approx {\mathcal E}$ 
(i.e., $\sigma \ll 1$) is realized for the poloidal field structure for which $B_p r^2$ 
continues decreasing along a field line, which is accompanied by decreasing $|B_{\varphi}| r$.
For the flux function we found, the magnetic flux tubes expand sideways rapidly, scaling as
$B_p r^2 \propto r^{-q}$ with $q>0$. In this case an arbitrarily efficient acceleration is available 
for an arbitrarily large $q$, as clearly demonstrated by \cite{fendt04,takahashi98}. 
Here $1/(B_p r^2)$ causes an effect similar to the cross section
of de Laval nozzle \cite{begelman94,komissarov10}, although 
the MHD equation of motion allows the flow to pass through 
the fast point even for the cases of a monotonic decrease of $B_p r^2$, as we have shown by 
deriving Eq. (\ref{eq:our_nozzle}). 

By using the wind solutions for the field line just inside the external boundary, we 
have calculated the external pressures $P_{\rm ext}$ as functions of $r$ (also of $z$ by using the
assumed form $z \propto r^{a_0}$) for various cases through the pressure balance condition
(Eq. \ref{eq:boundary}). Then we have found that $P_{\rm ext}$ has to decay very rapidly just beyond the 
light cylinder (by a factor of $\sim 10^5 - 10^6$ for $r_{\rm lc} \lesssim r \lesssim 3r_{\rm lc}$)
and asymptote to the power law decay $P_{\rm ext} \propto B_p^2 r^2$ for all the cases that we have 
calculated, i.e., the cases of the major energy conversion just beyond the fast point. 
Such external pressure profile might be due to thermal atmosphere or corona with 
closed magnetic field loops plus dilute wind above the thin disk, although the origin of 
such dilute winds is not so clear.

For the cases of $B_p r^2 \propto r^{-q}$, the acceleration $d\Gamma/dr$ 
monotonically decreases (as found from Eq. \ref{eq:approx_E}), and therefore a rapid
major energy conversion requires the initial rapid acceleration phase,
which is accompanied by the rapid decay of $P_{\rm ext}$. If $P_{\rm ext}$ maintains a rapid decay
even far beyond the light cylinder, $B_p r^2$ has to decrease with increasing $q$,
and the acceleration will be more rapid than the cases of constant $q$, although
the cases of variable $q$ are not included in the type of the flux function we have discussed.
If $P_{\rm ext}$ is constant beyond the region of the rapid decay region,
the shape of the boundary will be cylindrical (i.e., $r = {\rm const}$) at the region of the 
constant $P_{\rm ext}$, where $B_p$ and $\Gamma$ will be also constant. 
That is, the acceleration will stop beyond the region of the rapid pressure decay.

Komissarov et al. (2007, 2009) \cite{komissarov07,komissarov09} found global solutions 
above the inlet at $r < r_{\rm lc}$ with a fixed paraboloidal wall as the external boundary, 
by using numerical methods. They show that the poloidal fields near the jet axis are
self-collimated, leading to a decrease of $B_p r^2$ and an efficient fluid acceleration 
in the main body of the flow, and call this {\it collimation-acceleration mechanism}.
However, their solutions do not include the cases with rapid acceleration or 
rapid external pressure decay just beyond the light cylinder. 
Our results indicate that the rapid acceleration requires the rapid external pressure
decay, which might arise in the numerical calculations if the rigid wall suddenly expands 
near the light cylinder as shown in Figure \ref{fig:field_shape}. Such cases
can be seen in numerical calculations of the global structure by Tchekhovskoy et al.
(2010) and Komissarov et al. (2010) \cite{tchekhovskoy10,komissarov10}, 
in which the opening angle of the wall is set to suddenly
become larger at some large distance. The numerical results show that the poloidal
field line expands at the transition point and the rarefaction wave propagates inward, 
leading to additional rapid acceleration. Note that $\sigma \ll 1$ is realized near the wall.
The authors consider the context of
GRB jet breakouts from dense stellar envelope region into dilute interstellar 
medium, and called this {\it rarefaction-acceleration mechanism} \cite[see also][]{aloy06,mizuno08}. 
Thus our model may correspond to the rarefaction acceleration at the jet launching site.

A highly non-uniform distribution of the poloidal field strength at the inlet might
cause an additional effect for the energy conversion. For instance, if the field strength is 
much weaker near the jet axis, i.e., around the central BH, the flux tubes could expand towards
the jet axis. 

The field structure inside the light cylinder does not strongly affect the acceleration
efficiency, but may affect on the mass injection rate. If the angle between the field line 
and the accretion flow is small, the centrifugal force can easily blow 
the particles against the gravitational force, and 
the outflow becomes non-relativistic \cite{blandford82,spruit96}. 
We speculate that the shape of the field line nearly vertical above the accretion flow,
as illustrated in Figure \ref{fig:field_shape}, is effective to suppress mass injection from
the accretion flow and allow the outflow to be relativistic. Small amount of mass can be injected 
through electron-positron pair creation by collisions of two photons emitted from the
the corona \cite[][and references therein]{mckinney05} and/or through electron-proton 
creation by decays of neutrons escaped from the corona \cite{toma12}.

The rapid external pressure decay just beyond the light cylinder may arise more easily for 
geometrically thin accretion flows.
For geometrically thick accretion flows, such as advection-dominated accretion
flows \cite[e.g.,][]{narayan94,mckinney07}, the outflow region will be confined by
the dense accretion flow (probably with disk wind), so that the rarefaction acceleration 
seems difficult.
The central region of GRB jets within collapsing stars are expected to have high pressure
\cite[e.g.,][]{sekiguchi11}. 
Thus the rarefaction acceleration may not be applicable for them, although such acceleration 
is effective at the jet breakout sites from the dense region into dilute interstellar medium, 
as discussed above. For compact stars mergers, the density can be very low above 
accretion flow if a BH is formed \cite[e.g.,][]{hotokezaka12}, so that the rarefaction
acceleration at the jet launching site could be effective \cite{aloy06}.

\section*{Acknowledgment}
We thank the referee for useful comments.
K. T. thanks Susumu Inoue, Kunihito Ioka, Hideo Kodama, Kazunori Kohri, Koutarou Kyutoku, Akira Mizuta, 
Shin-ya Nitta, Takeru K. Suzuki, and Hajime Takami for useful discussions.
This work is partly supported by JSPS Research Fellowships for Young Scientists 
No. 231446.

\appendix

\section{Derivation of Eq. (\ref{eq:our_nozzle})}

We derive Eq. (\ref{eq:our_nozzle}) by taking an asymptotic limit of Eq. (\ref{eq:eom_l}), which
can be rewritten as
\begin{equation}
u_p du_p = \frac{u_{\varphi}^2}{r}dr - \frac{B_{\varphi}}{4\pi \rho c^2 r}d(r B_{\varphi}).
\label{eq:eom_l_app}
\end{equation}
Here all the derivatives represent those in the direction of ${\mathbf B}_p$. 

First, we have from Eq. (\ref{eq:diff_E})
\begin{equation}
d\Gamma = \frac{{\mathcal E}-\Gamma}{-rB_{\varphi}} d(r B_{\varphi}),
\end{equation}
and by differentiating Eq. (\ref{eq:Omega}) with $d\Omega=0$
\begin{equation}
v_{\varphi} \frac{dr}{r^2} = \frac{1}{r}dv_{\varphi} - \frac{r B_{\varphi}}{\mathcal S} dv_p
-\frac{v_p}{\mathcal S}d(r B_{\varphi}) + \frac{v_p r B_{\varphi}}{{\mathcal S}^2} d{\mathcal S}.
\end{equation}
Combining these two equations with Eq. (\ref{eq:eom_l_app}), we obtain
\begin{eqnarray}
\left[\left(1+2\frac{u_p B_p}{u_{\varphi}B_{\varphi}}\right)\frac{1}{u_p^2}
+ \frac{u_p^2 -1}{u_{\varphi}u_p} \frac{B_p}{B_{\varphi}}
- \left(1+\frac{B_{\varphi}B_p}{4\pi \rho c^2 u_{\varphi} u_p}\right) \frac{\Gamma}{{\mathcal E}-\Gamma}\right]
\frac{d\Gamma}{\Gamma} \nonumber \\
+ \left(1+2\frac{u_p B_p}{u_{\varphi}B_{\varphi}}\right) 
\frac{(-v_{\varphi}dv_{\varphi})}{v_p^2} = \frac{d{\mathcal S}}{\mathcal S},
\end{eqnarray}
where we have used the relation $v_p dv_p + v_{\varphi}dv_{\varphi} = c^2d\Gamma/\Gamma^3$.

The parenthesis in the first term can be rewritten by using ${\mathcal E}-\Gamma
= -r\Omega B_{\varphi} B_p/(4\pi \rho u_p c^3)$ as
\begin{eqnarray}
[\; ] &=& \left[ \Gamma - \frac{B_{\varphi}^2}{4\pi \rho c^2 \Gamma} \frac{r\Omega}{r\Omega - v_{\varphi}}
\left(\frac{1}{u_p^2} + \frac{c^2}{r\Omega v_{\varphi}}\right) 
+ \frac{B_p^2 r\Omega}{4\pi \rho u_{\varphi}} \left(1+\frac{1}{u_p^2}\right)
\right] \frac{-1}{{\mathcal E}-\Gamma} \nonumber \\
&=& \left[ u_p^2 - \frac{B_{\varphi}^2}{4\pi \rho c^2} 
\frac{v_p^2/c^2}{1-\frac{c^2}{r^2 \Omega^2}(1-\frac{\Gamma_{\rm in}}{\Gamma})}
\left(\frac{1}{u_p^2} + \frac{1}{1-\frac{\Gamma_{\rm in}}{\Gamma}}\right) \right. \nonumber \\
&&~~~~ \left. + \frac{B_p^2}{4\pi \rho c^2} \frac{r^2 \Omega^2}{c^2} \frac{v_p^2/c^2}{1-\frac{\Gamma_{\rm in}}{\Gamma}}
\left(1+\frac{1}{u_p^2}\right)
\right] \frac{-1}{({\mathcal E}-\Gamma)\Gamma v_p^2/c^2},
\end{eqnarray}
where we have used Eq. (\ref{eq:v_phi}). By making $u_f^2$ in the parenthesis, the whole equation reduces to
\begin{eqnarray}
\left\{ u_p^2 - u_f^2 + \frac{B_{\varphi}^2}{4\pi \rho c^2} \left[1- 
\frac{v_p^2/c^2}{1-\frac{c^2}{r^2\Omega^2}(1-\frac{\Gamma_{\rm in}}{\Gamma})} 
\left(\frac{1}{u_p^2} + \frac{1}{1-\frac{\Gamma_{\rm in}}{\Gamma}} \right) \right] \right. \nonumber \\
\left. - \frac{B_p^2}{4\pi \rho c^2} \frac{r^2 \Omega^2}{c^2} \left[1-
\frac{v_p^2/c^2}{1-\frac{\Gamma_{\rm in}}{\Gamma}} \left(1+\frac{1}{u_p^2}\right) \right]
+\frac{B_p^2}{4\pi \rho c^2} \right\}
\frac{d\Gamma}{u_p^2({\mathcal E}-\Gamma)} \nonumber \\
-\left(1+2\frac{u_p B_p}{u_{\varphi} B_{\varphi}}\right) \frac{(-v_{\varphi}dv_{\varphi})}{v_p^2}
= \frac{-d{\mathcal S}}{{\mathcal S}}.
\end{eqnarray}
This equation can be approximated into Eq. (\ref{eq:our_nozzle}) by using Eqs. (\ref{eq:app_Bphi}) and (\ref{eq:app_vp})
and neglecting terms of orders of $c^2/r^2\Omega^2 \Gamma^2$, $c^4/r^4 \Omega^4$, and $1/\Gamma^4$
in the parenthesis $\{\; \}$.

\end{document}